\documentclass[a4paper,twocolumn,english,aps,prb,amsmath,showpacs,amssymb]{revtex4-1}
\usepackage[pdftex]{graphicx,hyperref}
\usepackage{verbatim}
\usepackage{amsbsy}
\usepackage{amsmath}
\usepackage{color}
\usepackage{bm}
\usepackage{marvosym}
\usepackage{wasysym}

\begin{document}

\title{Intervalley scattering of graphene massless Dirac fermions at 3-periodic grain boundaries}

\author{J. N. B. Rodrigues}

\affiliation{Centre for Advanced 2D Materials, National
  University of Singapore, 6 Science Drive 2, Singapore 117546}

\pacs{81.05.ue, 72.80.Vp}

\date{\today}

\begin{abstract}
  We study how low-energy charge carriers scatter off periodic and
  linear graphene grain boundaries oriented along the zigzag direction
  with a periodicity three times greater than that of pristine
  graphene. These defects map the two Dirac points into the same
  position, and thus allow for intervalley scattering to
  occur. Starting from graphene's first-neighbor tight-binding model
  we show how can we compute the boundary condition seen by graphene's
  massless Dirac fermions at such grain boundaries. We illustrate this
  procedure for the 3-periodic pentagon-only grain boundary, and then
  work out the low-energy electronic scattering off this linear
  defect. We also compute the effective generalized potential seen by
  the Dirac fermions at the grain boundary region.
\end{abstract}

\maketitle

\section{Introduction}

Chemical vapor deposition (CVD) of graphene on metal 
surfaces\cite{Li_Science:2009,Reina_NanoLett:2009,Kim_Nature:2009,Bae_NatureNanotech:2010} is currently viewed as 
one of the most promising scalable methods for economically producing large and abundant high-quality monolayer 
graphene sheets. It is thus greatly important to fully understand and control the behavior of electrons on this 
form of graphene. 

CVD graphene, as any other solid grown by chemical vapor deposition, is generally a polycrystal composed by several 
grains with distinct crystallographic orientations. These grains are separated by grain boundaries 
(GBs),\cite{Meyer_NL:2008,Lahiri_NatureNanotech:2010,Huang_Nature:2011,Kim_ACSNano:2011,Incze_APL:2011,Colin_PRB:2015} 
which due to the $sp^{2}$ bonding structure of carbon atoms in graphene, are typically made of pentagonal, heptagonal and 
octagonal rings of carbon atoms.\cite{Meyer_NL:2008,Lahiri_NatureNanotech:2010,Huang_Nature:2011,Kim_ACSNano:2011,Colin_PRB:2015} 
Grain boundaries generally intercept each other at random angles, being neither periodic nor perfect straight lines. 

The properties of CVD graphene flakes are strongly influenced by the quantity, distribution and microscopic character
of its grain boundaries.\cite{Yazyev_NatNano:2014,Cummings_AdvMat:2014} Each type of grain boundary exhibits distinctive 
chemical,\cite{Yasaei_NatComm:2014,Seifert_2DMat:2015} mechanical\cite{Grantab_Science:2011,Lee_Science:2013} and 
electronic\cite{Yu_NatureMat:2011,Jauregui_SSComm:2011,Tsen_Science:2012} properties.

This is particularly evident in what concerns the electronic transport in CVD graphene. For instance, there is abundant
experimental evidence that the details of the CVD-growth recipes used to synthesize CVD graphene flakes greatly constrain 
its transport properties.\cite{Reina_NanoLett:2009,Li_Science:2009,Huang_Nature:2011,Bae_NatureNanotech:2010,Kochat_NanoLett:2016} 
Furthermore, application of strain and chemical decoration are also expected to strongly influence CVD-graphene transport
properties.\cite{Zhang_JPCC:2014,Nguyen_Nanoscale:2016} In particular, as shown by several recent experiments probing the 
transport properties of single grain boundaries,\cite{Yu_NatureMat:2011,Jauregui_SSComm:2011,Tsen_Science:2012,Yasaei_NatComm:2014} 
the electron-scattering off a grain boundary is determined by the its microscopic details and the relative orientation of the 
grains it separates.\cite{Yazyev_NatureMat:2010}

Observation and probing of graphene grain boundaries has been constantly refined in recent years, as shown by a quick
survey of the recent literature in the field.\cite{Kochat_NanoLett:2016,Kim_SciRep:2015,Gunlycke_PRB:2015,Ago_AcsnNano:2016} 
More interestingly, several promising new methods of controlling 
and manipulating the position, orientation and microscopic configuration of grain boundaries have been recently 
unveiled.\cite{Song_NanoLett:2011,Kurasch_NanoLett:2012,Chen_PRB:2014,Yang_JACS:2014} Some of these methods
allow for the creation of periodic and straight grain boundaries,\cite{Kurasch_NanoLett:2012,Chen_PRB:2014,Yang_JACS:2014}
whose transport properties have been extensively investigated theoretically.\cite{Gunlycke_PRL:2011,Jiang_PLA:2011,Liwei_PRB:2012,Rodrigues_GBs-CA_PRB:2012,Rodrigues_GBs-TB_JPCM:2013,Ebert_JPCM:2014,Paez_PRB:2015} 
This widens the prospects for the engineering of graphene-based electronic devices that take advantage of the scattering
properties of these grain boundaries, to manipulate graphene electrons' various degrees of freedom, such as its valley quantum
number.

Following these recent advances, in this manuscript, we will focus our attention on the electronic properties of a particular 
class of periodic and linear grain boundaries that is often disregarded in the literature. Namely, we will investigate grain 
boundaries with periodicities such that both Dirac points (on each side of the GB) are mapped into the $\Gamma$ point of 
the projected Brillouin zone -- see Ref. \onlinecite{Yazyev_NatureMat:2010} for a brief discussion of their properties. Due to 
this mapping of the Dirac points, such grain boundaries allow for intervalley scattering of low-energy charge carriers. In what
follows we will show how can we work out the low-energy electronic scattering off such GBs, and will see how the intervalley 
scattering depends on the system's microscopic details.

To keep things simple, we have chosen to investigate zigzag aligned linear grain boundaries separating two grains with the same 
orientation (also referred to in the literature as degenerate or zero misorientation-angle grain boundaries). Several such 
GBs were proposed in the context of ab-initio works both on graphene and on boron nitride: the t7t5 grain 
boundary,\cite{Botello-Mendez_Nanoscale:2011} the 7557 grain boundary\cite{Ansari_PCCP:2014} and the 8484 grain 
boundary.\cite{Han_Nanotechnology:2014} Their transport properties have been recently investigated under the
perspective of the tight-binding model.\cite{Paez_PRB:2015}

\begin{figure}
  \includegraphics[width=0.98\columnwidth]{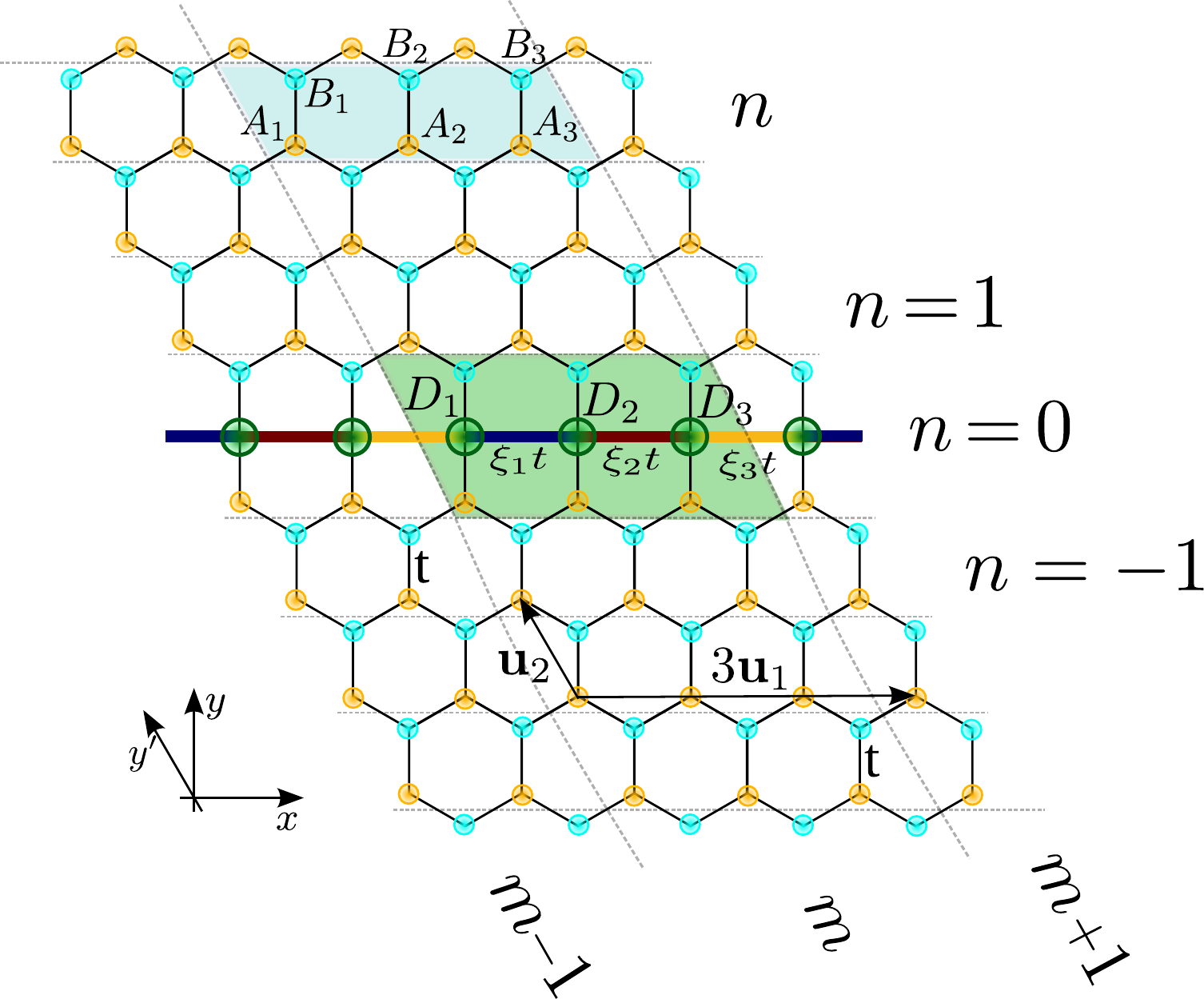}
  \caption{Scheme of a 3-periodic pentagon-only grain boundary. Notice
    that its unit cell has a periodicity of $3 \mathbf{u}_{1}$ along
    the $x$-direction. The lattice vectors used are $3 \mathbf{u}_{1}
    = 3 a (1,0)$ and $\mathbf{u}_{2} = (-1,\sqrt{3}) a/2$. The hopping
    parameters at the defect alternate in value reading (in units of
    $t$): $\xi_{1}$, $\xi_{2}$ and $\xi_{3}$.}
  \label{fig:Scheme-3FPentOnly}
\end{figure}

\begin{figure}
  \includegraphics[width=0.98\columnwidth]{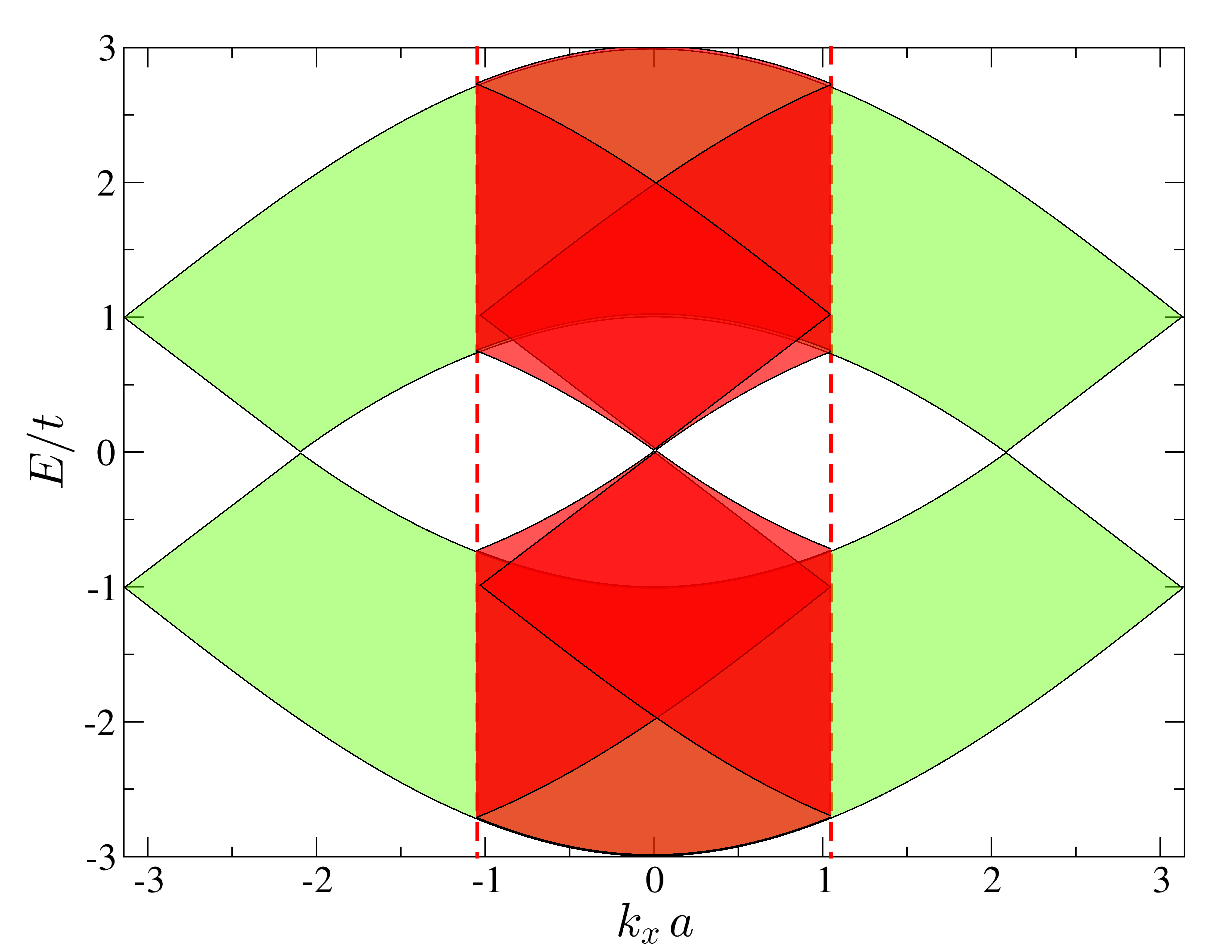}
  \caption{Pristine (and infinite) graphene energy spectrum projected
    along the $k_{y}$ direction (perpendicular to the defect). In
    green, the spectrum of pristine graphene (with the lattice vector
    along the zigzag direction given by $\mathbf{u}_{1}$). Its Dirac
    points are located at $k_{x} a = \pm 2 \pi / 3$. In red, the
    folded spectrum arising from choosing the lattice vector along the
    zigzag direction to be $3 \mathbf{u}_{1}$. In this case, we can
    clearly see that the two Dirac points are mapped into $k_{x} a =
    0$.}
  \label{fig:FBZ3-projected}
\end{figure}

In what follows we will concentrate on studying a simplistic but illustrative grain boundary representative of the 
above class. We will consider a pentagon-only like 
grain boundary\cite{Rodrigues_GBs-CA_PRB:2012,Rodrigues_GBs-TB_JPCM:2013} with a periodicity three times greater than that 
of pristine graphene (along the zigzag direction -- see Fig. \ref{fig:Scheme-3FPentOnly}). As desired, such a periodicity 
ensures that the Dirac points are mapped into the $\Gamma$ point as discussed above -- see Fig. \ref{fig:FBZ3-projected}. 
In the remaining of this text we will call this GB by {\it 3-periodic pentagon-only} grain boundary. 

This particular grain boundary must be seen as a minimal model representing the general class of zigzag aligned 
3-periodic grain boundaries (such as the t7t5, the 7557 and the 8484 grain 
boundaries\cite{Botello-Mendez_Nanoscale:2011,Ansari_PCCP:2014,Han_Nanotechnology:2014}). Synthesis of grain 
boundaries in this class may be facilitated by CVD deposition of graphene on poly-crystalline substrates with 
linear grain boundaries with the appropriate periodicity (i.e. $\approx 7.4 \mathring{A}$). Decoration of other 
grain boundaries [e.g. the $zz(558)$] with periodic arrays of molecules may also give rise to 3-fold periodic 
grain boundaries that allow for intervalley scattering of low-energy charge carriers.
 
As is well known, graphene's low energy charge carriers behave as massless Dirac fermions. These are governed 
by a Hamiltonian composed of two copies of the 2D Dirac Hamiltonian, each one of them valid around one of the 
two Dirac points.\cite{Semenoff_PRL:1984} In this limit the grain boundary essentially acts as a one-dimensional 
line that imposes a boundary condition on the Dirac spinors living on the semi-planes above and bellow the grain
boundary. Such boundary condition will result in a discontinuity in the spinors across the defect and will 
control its scattering properties.\cite{Liwei_PRB:2012,Rodrigues_GBs-CA_PRB:2012} In alternative, the grain boundary can also 
be thought of as a finite width strip containing a generalized potential that constrains the dynamics of the 
massless Dirac fermions.\cite{Rodrigues_GBs-CA_PRB:2012,Rodrigues_GBs-TB_JPCM:2013,Ebert_JPCM:2014}

The specific form of the boundary condition seen by the massless Dirac fermions at the 3-periodic pentagon-only 
grain boundary is determined by the details of its microscopic tight-binding model. In calculating it, we will
follow the methodology developed for the cases of the pentagon-only, $zz(558)$ and $zz(5757)$ grain 
boundaries.\cite{Rodrigues_GBs-CA_PRB:2012,Rodrigues_GBs-TB_JPCM:2013} Below, we show that the boundary condition 
obtained from the tight-binding gives rise, in the low-energy limit, to a boundary condition explicitly introducing 
intervalley scattering. Furthermore, we will show how, starting from the grain boundary's microscopic details, can we 
determine the generalized potential associated with viewing the grain boundary as a finite width strip with a
potential constraining the Dirac fermions' dynamics.

Before proceeding, we detail the structure of this text. In Section \ref{sec:LowEn-3Periodic} we discuss general 
properties of Dirac fermion scattering off 3-periodic grain boundaries. In Section \ref{sec:3PO} we solve the electronic 
scattering off a 3-periodic pentagon-only grain boundary: we start by computing the tight-binding boundary condition 
matrix relating electronic amplitudes on each side of the grain boundary (see sub-Section \ref{sec:TBmodel-3PO}); we
then derive the boundary condition matrix {\it seen} by the low-energy charge carriers (i.e. by the massless Dirac 
fermions) at the grain boundary (see sub-Section \ref{sec:CA-atGB}); finally, in sub-Section \ref{sec:Transmittance}
we compute the transmission probabilities for different choices of the microscopic hopping parameters at the grain
boundary. We close with Section \ref{sec:conclusion} where we overview the main results of the manuscript.

\section{General properties of low-energy electron transport across a 3-periodic grain boundary}

\label{sec:LowEn-3Periodic}

Let us start by considering the case of a general 3-periodic grain boundary, i.e. a grain boundary with a periodicity (and 
orientation) defined by the vector $\mathbf{R} = n \mathbf{u}_{1} + m \mathbf{u}_{2}$, where $n, m \in \mathbb{N}$ and 
$n+m$ is a multiple of $3$ -- see Fig. \ref{fig:Scheme-3FPentOnly} where $n=3$ and $m=0$. The presence of such a grain
boundary in graphene, breaks the translation symmetry along the direction perpendicular to the grain boundary. Furthermore,
the periodicity of these grain boundaries happens to fold the first Brillouin zone in such a way that the two Dirac points 
are mapped into the $\Gamma$-point of the projected Brillouin zone. It is thus natural to expect that intervalley scattering 
off these nanostructures is generally allowed at low energies.

As a consequence, when addressing the problem of electronic scattering across such kind of grain boundaries,
we need to consider both valleys. Instead of working with two separate copies
of the Dirac Hamiltonian $H_{\nu} = v_{F} \big( \nu \sigma_{1} \, \partial_{x} + \sigma_{2} \, \partial_{y} 
\big)$ valid in the vicinity of each Dirac point $\mathbf{K}_{\nu=\pm}$, we have to work with both
copies simultaneously, i.e. with the Hamiltonian
\begin{eqnarray}
  H &=& v_{F} \big( \tau_{3} \otimes \sigma_{1} \, \partial_{x} + \tau_{0} \otimes \sigma_{2} \, 
  \partial_{y} \big) \,, \label{eq:H2K}
\end{eqnarray}
where $\tau_{i}$ and $\sigma_{i}$ ($i=1,2,3$) stand for the $2 \times 2$ Pauli matrices acting, respectively,
on the valley and pseudo-spin degrees of freedom, while $\tau_{0}$ and $\sigma_{0}$ stand for the $2 \times 2$ 
identity matrix. In the above equation (and in the remaining of this text) we have set $\hbar = 1$. As 
we are considering a Hamiltonian that is independent of (real) spin, we will always omit the spin degree of freedom.

The presence of a periodic grain boundary in a graphene 
flake imposes a discontinuity between the Dirac fermion's spinors on each of the grain boundary's 
sides.\cite{Liwei_PRB:2012,Rodrigues_GBs-CA_PRB:2012,Rodrigues_GBs-TB_JPCM:2013,Ebert_JPCM:2014}
However, grain boundaries that are $3$-periodic also connect the two valleys through non-zero intervalley scattering 
matrix elements. For simplicity, let us consider the case of a zigzag-aligned (i.e. $x$-aligned) $3$-periodic
grain boundary located at $y=0$. Such a grain boundary imposes the following general boundary condition on the 
Dirac spinors:
\begin{eqnarray}
  \Psi(x,0^{+}) &=& \mathcal{M} \Psi(x,0^{-}) \,, \label{eq:GeneralBC-M}
\end{eqnarray}
where $\Psi(\mathbf{r}) = [\psi_{+}(\mathbf{r}),\psi_{-}(\mathbf{r})]^{T}$ are $4$-spinors since
$\psi_{\pm} = [\varphi_{a \pm}(\mathbf{r}),\varphi_{b \pm}(\mathbf{r})]^{T}$ stands for the 2-spinor 
describing Dirac fermions living in the $\mathbf{K}_{\pm}$ valley. The $4 \times 4$ matrix $\mathcal{M}$
can be written in the general form
\begin{eqnarray}
  \mathcal{M} &=& \left [ \begin{array}{cc} \mathcal{M}_{++} & \mathcal{M}_{+-} \\ \mathcal{M}_{-+} 
      & \mathcal{M}_{--} \end{array} \right ] \,, \label{eq:GenMca}
\end{eqnarray}
where $\mathcal{M}_{\pm \pm}$ ($\mathcal{M}_{\pm \mp}$) are $2 \times 2$ matrices controlling the valley
preserving (intervalley) scattering across the grain boundary.

The matrix $\mathcal{M}$ must satisfy the flux conservation condition $\mathcal{M}^{\dagger} J_{y} \mathcal{M} 
= J_{y}$, where $J_{y} = \tau_{0} \otimes \sigma_{y}$ stands for the conserved current along the direction 
perpendicular to the grain boundary. This stems from the hermiticity of the tight-binding Hamiltonian (that
enforces current conservation at the GB in the tight-binding model). Furthermore, whenever we deal with 
non-magnetic grain boundaries, the boundary condition must as well be time-reversal invariant $\mathcal{T}^{-1} 
\mathcal{M} \mathcal{T} = \mathcal{M}$. Recall that the time-reversal operation exchanges the two Dirac cones 
and applies complex conjugation, $\mathcal{T} = \tau_{1} \otimes \sigma_{0} \, \mathcal{C}$.

A Dirac fermion from the $\mathbf{K}_{\nu}$ valley ($\nu=\pm1$), incoming from $y=-\infty$ (see Fig. 
\ref{fig:Scheme-3FPentOnly}) will be partially transmitted and partially reflected at the grain boundary. In 
the absence of a potential difference between the two sides of the grain boundary, the wave function reads
\begin{subequations} \label{eq:WaveFunctions_Dirac-1}
\begin{eqnarray} 
  \Psi^{L}(\mathbf{r}) &=& \Psi_{\mathbf{q}, s}^{\nu, >}(\mathbf{r}) + \rho_{\nu,\nu} \, 
  \Psi_{\mathbf{q}, s}^{\nu, <}(\mathbf{r}) + \rho_{-\nu,\nu} \, \Psi_{\mathbf{q}, s}^{-\nu, <}(\mathbf{r}) \,,
\end{eqnarray}
on the lower half-plane (i.e. $y<0$), while for the upper half-plane (i.e. $y>0$) we have
\begin{eqnarray}
  \Psi^{U}(\mathbf{r}) &=& \tau_{\nu,\nu} \, \Psi_{\mathbf{q}, s}^{\nu, >}(\mathbf{r}) 
  + \tau_{-\nu,\nu} \, \Psi_{\mathbf{q},s}^{-\nu, >}(\mathbf{r}) \,.
\end{eqnarray}
\end{subequations}
In Eqs. (\ref{eq:WaveFunctions_Dirac-1}) the $\Psi_{\mathbf{q},s}^{\nu,\lessgtr}(\mathbf{r})$ are 4-spinors which read
\begin{subequations} \label{eq:WaveFunctions_Dirac-2} 
  \begin{eqnarray}
    \Psi_{\mathbf{q} s}^{+,>}(\mathbf{r}) &=& \frac{1}{\sqrt{2}} \left[\begin{array}{cccc} s \, e^{-i\theta_{\mathbf{q}}^{+}}, & 1, & 0, & 0
      \end{array}\right]^{T} e^{i(q_{x}x+q_{y}y)} \,, \\
    \Psi_{\mathbf{q} s}^{-,>}(\mathbf{r}) &=& \frac{1}{\sqrt{2}} \left[\begin{array}{cccc} 0, & 0, & s \, e^{-i\theta_{\mathbf{q}}^{-}}, & 1
      \end{array}\right]^{T} e^{i(q_{x}x+q_{y}y)} \,, \\
    \Psi_{\mathbf{q} s}^{+,<}(\mathbf{r}) &=& \frac{1}{\sqrt{2}} \left[\begin{array}{cccc} s \, e^{-i \bar{\theta}_{\mathbf{q}}^{+}}, & 1, & 0, & 0
      \end{array}\right]^{T} e^{i(q_{x}x-q_{y}y)} \,, \\
    \Psi_{\mathbf{q} s}^{-,<}(\mathbf{r}) &=& \frac{1}{\sqrt{2}} \left[\begin{array}{cccc} 0, & 0, & s \, e^{-i \bar{\theta}_{\mathbf{q}}^{-}}, & 1
      \end{array}\right]^{T} e^{i(q_{x}x-q_{y}y)} \,.
  \end{eqnarray}
\end{subequations}
In the above expressions $\mathbf{r}=(x,y)$, $\mathbf{q}=(q_{x},q_{y})$, while $\theta_{\mathbf{q}}^{\nu}$ and $\bar{\theta}_{\mathbf{q}}^{\nu}
= - \theta_{\mathbf{q}}^{\nu}$ stand for the complex phases of, respectively, $\nu q_{x} + i q_{y}$ and $\nu q_{x} - i q_{y}$. Furthermore,
$s$ stands for the sign of the energy, distinguishing electrons and and holes, while $\rho_{\pm,\pm}$ ($\rho_{\pm,\mp}$) and $\tau_{\pm,\pm}$
($\tau_{\pm,\mp}$) respectively stand for the valley preserving (intervalley) reflection and transmission coefficients.

Since the conserved current associated with a given propagating mode is the same for all modes, then the four transmission 
and reflection probabilities are simply given by $T_{\nu \eta} = \vert \tau_{\nu \eta} \vert^{2}$ and $R_{\nu \eta} = \vert 
\rho_{\nu \eta} \vert^{2}$, for $\nu=\pm1$ and $\eta=\pm1$. The transmission and reflection coefficients are obtained by 
solving the system of linear equations originating from imposing the boundary condition Eq. (\ref{eq:GeneralBC-M}) on
the wave function written in Eqs. (\ref{eq:WaveFunctions_Dirac-1}) and (\ref{eq:WaveFunctions_Dirac-2}). 
Therefore, the transmission and reflection probabilities will not only depend on the angle of incidence of the electron 
into the GB, but will also (strongly) depend on the microscopic properties of the grain boundary through the matrix 
elements of $\mathcal{M}$. Below, following Refs. \onlinecite{Rodrigues_GBs-CA_PRB:2012,Rodrigues_GBs-TB_JPCM:2013}, we 
will show how can we compute the boundary condition matrix $\mathcal{M}$ from the tight-binding model of the grain boundary.

\begin{figure}
  \includegraphics[width=0.99\columnwidth]{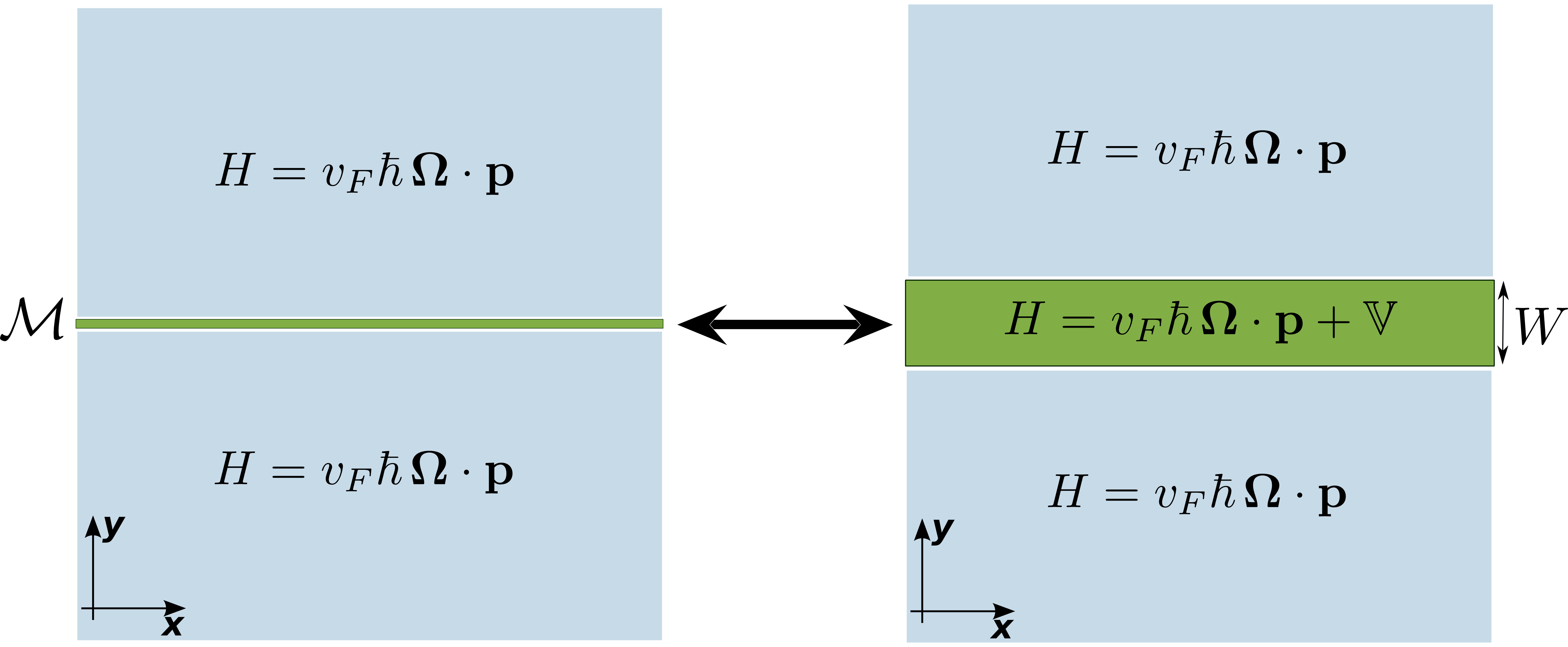}
  \caption{Schematic representation of the two perspectives used to
    analyze how do graphene's low-energy charge carriers scatter off
    a grain boundary. On the left-hand-side, the massless Dirac
    fermions are constrained by
    $\Psi(x,0^{+})=\mathcal{M}\Psi(x,0^{-})$, which gives rise to a
    discontinuity of the Dirac spinors at the line separating the two
    half-planes [where the Dirac Hamiltonian governs the physics: $H =
    v_{F} \, \boldsymbol{\Omega} \cdot \mathbf{p}$, with
    $\boldsymbol{\Omega} \equiv
    (\tau_{3}\otimes\sigma_{1},\tau_{0}\otimes\sigma_{2})$]. On the
    right-hand-side, these two half-planes are separated by a finite
    width strip where the Dirac fermions are subject to a generalized
    potential $\mathbb{V}$.}
  \label{fig:two-perspectives}
\end{figure}

However, before proceeding we will briefly discuss an often useful alternative viewpoint for such scattering problems
(see Ref. \onlinecite{Rodrigues_GBs-CA_PRB:2012,Rodrigues_GBs-TB_JPCM:2013}). Instead of considering that in
the low-energy limit the grain boundary simply imposes a discontinuity on the massless Dirac fermions' spinors along 
a line parallel to the $x$-axis, we will consider that the grain boundary can be viewed as a finite strip of width 
$W$ that exists in $\vert y\vert < W/2$ and extends along the $x$ direction -- see right-hand-side of Fig. 
\ref{fig:two-perspectives}. On each side of this strip the Dirac fermions will be governed by Eq. (\ref{eq:H2K}), 
while inside the strip there will also be a general local potential of the form 
\begin{eqnarray}
  \mathbb{V} &=&  \sum_{\alpha, \beta = 0}^{3} V_{\alpha \beta} \, \tau_{\alpha} \otimes \sigma_{\beta} \,. \label{eq:GenPotV}
\end{eqnarray}

The equivalence between these two viewpoints becomes obvious when we integrate out the Dirac equation in the finite width 
strip (i.e. between $-W/2 < y < W/2$) while satisfying the constrain $W \times V_{\alpha \beta} \to v_{\alpha \beta}$ as 
$W\to0$. As we show in detail in Appendix \ref{app:GenPot}, this integration gives rise to the following boundary 
condition matrix
\begin{eqnarray}
  \mathcal{M} &=& e^{{-i \frac{\tau_{0} \otimes \sigma_{2}}{v_{F}} \mathbb{V}}} \,,
\end{eqnarray}
which allows us to connect the matrix elements of $\mathcal{M}$ (determined from a tight-binding microscopic model) and 
the effective generalized local potential felt by the Dirac fermions inside the finite width strip. 

Within this perspective, there are two interfaces at which we must ensure the continuity of the wave function, namely 
$\Psi(x,-w^{-}) = \Psi(x,-w^{+})$ and $\Psi(x,w^{-}) = \Psi(x,w^{+})$ where $w=W/2$. These
two equalities correspond to eight conditions ($\Psi$ are 4-spinors) that will determine the eight unknown scattering 
coefficients (region inside the strip requires four additional coefficients).

\section{The 3-periodic pentagon-only grain boundary}

\label{sec:3PO}

In what follows we will make the statements of the previous section concrete by investigating the
electronic transport across the 3-periodic pentagon-only grain boundary (see Fig. \ref{fig:Scheme-3FPentOnly}).
We will start by briefly sketching how can we compute the tight-binding boundary condition matrix relating the
wave function above and below the grain boundary\cite{Rodrigues_GBs-TB_JPCM:2013,Paez_PRB:2015} -- see sub-Section 
\ref{sec:TBmodel-3PO}. From that result we will then compute the boundary condition matrix seen by the massless
Dirac fermions at the grain boundary -- see sub-Section \ref{sec:CA-atGB}. Finally, in sub-Section 
\ref{sec:Transmittance} we will work out the scattering problem and analyze the valley preserving and intervalley 
transmittance for specific sets of microscopic parameters defining the 3-periodic pentagon-only grain boundary.

\subsection{The tight-binding model for the grain boundary}

\label{sec:TBmodel-3PO}

Consider a first-neighbor tight-binding model for electrons in the $p_{\textrm{z}}$-orbitals of graphene, where we 
define the pristine honeycomb direct lattice vectors as (see Fig. \ref{fig:Scheme-3FPentOnly}): $\mathbf{u}_{1} 
= a (1,0)$ and $\mathbf{u}_{2} = (-1,\sqrt{3}) a/2$. As we want to study a zigzag-oriented grain boundary with 
periodicity $3 \mathbf{u}_{1}$, we choose a {\it bulk} unit cell defined by the lattice 
vectors $3 \mathbf{u}_{1}$ and $\mathbf{u}_{2}$ as sketched in Fig. \ref{fig:Scheme-3FPentOnly}. Fourier transforming 
along the $3 \mathbf{u}_{1}$ direction diagonalizes the system's Hamiltonian with respect to the variable $m$, 
introducing the quantum number $k_{x}$. The corresponding bulk tight-binding equations can then be written as
\begin{subequations} \label{eq:BulkTBeqs}
  \begin{eqnarray}
    -\frac{\epsilon}{t}\mathbf{A}(n) &=& W_{A}^{\dagger}\mathbf{B}(n-1)+\mathbf{B}(n) \,,\\
    -\frac{\epsilon}{t}\mathbf{B}(n) &=& \mathbf{A}(n)+W_{A}\mathbf{A}(n+1) \,,
  \end{eqnarray}
\end{subequations}
where $\mathbf{Z}(n)=[Z_{1}(n),Z_{2}(n),Z_{3}(n)]^{T}$ for $Z=A,B$ and the matrix $W_{A}$ is defined in Eq. 
(\ref{eq:WAmatrix}).

Eqs. (\ref{eq:BulkTBeqs}) can be condensed in the form of a transfer matrix 
equation\cite{Rodrigues_GBs-CA_PRB:2012,Rodrigues_GBs-TB_JPCM:2013} relating amplitudes at the atoms 
of the unit cell located at $\mathbf{r} = (n-1) \mathbf{u}_{2}$ with the amplitudes at the atoms of the unit cell
located at $\mathbf{r} = n \mathbf{u}_{2}$. Such an equation reads
\begin{eqnarray}
  \mathbf{L}(n) &=& \mathbb{T}(\epsilon,k_{x}) \mathbf{L}(n-1) \,, \label{eq:transfMatEq}
\end{eqnarray}
with $\mathbf{L}(n) = [A_{1}(n),B_{1}(n),A_{2}(n),B_{2}(n),A_{3}(n),B_{3}(n)]^{T}$, while the transfer matrix
$\mathbb{T}(\epsilon,k_{x})$ is given by
\begin{eqnarray}
  \mathbb{T}(\epsilon,k_{x}) &=& R \, \mathbb{Q}_{1} \, \mathbb{Q}_{2} \, R^{T} \,. \label{eq:TransferMatrixCoup}
\end{eqnarray}
In the above equation, matrix $R$ is simply used to change from the basis $\{ B_{1},B_{2},B_{3},A_{1},A_{2} ,A_{3} \}$ 
into the basis $\{A_{1},B_{1},A_{2},B_{2},A_{3},B_{3} \}$. This matrix is written in Eq. (\ref{eq:Rmatrix}), while matrices
$\mathbb{Q}_{1}$ and $\mathbb{Q}_{2}$ are written in Eqs. (\ref{eq:Q1Q2}).

As was shown in Ref. \onlinecite{Paez_PRB:2015}, we can employ a basis transformation $\widetilde{\mathbf{L}}(n) = 
\Lambda(k_{x}) \mathbf{L}(n)$ that makes the transfer matrix block diagonal, $\widetilde{\mathbb{T}} = 
\textrm{diag}[ \mathbb{T}_{h}, \mathbb{T}_{l-}, \mathbb{T}_{l+}]$. We will use the notation $\widetilde{\mathbf{L}}(n)
= \{ A_{h}, B_{h}, A_{l-}, B_{l-}, A_{l+}, B_{l+} \}$ to identify the elements of a vector in this basis. The three matrices
$\mathbb{T}_{h}$, $\mathbb{T}_{l-}$ and $\mathbb{T}_{l+}$ on the diagonal of $\widetilde{\mathbb{T}}$ are $2 \times 2$ 
matrices that depend on both $\epsilon$ and $k_{x} a$. They are written in Eq. (\ref{eq:TMdiags}), while the matrix 
$\Lambda(k_{x})$ is written in Eq. (\ref{eq:LambdaMat}). Each one of the $\mathbb{T}_{h}$, $\mathbb{T}_{l-}$ and 
$\mathbb{T}_{l+}$ matrices correspond to one of the three propagation modes of the problem. Around $k_{x} a = 0$ two of 
these modes are low-energy (corresponding to each of the two Dirac cones), while the other one is a high-energy mode 
-- see Appendix \ref{app:BulkTB}.

In a similar way, we can compute the boundary condition matrix that relates electronic amplitudes below and above 
the 3-periodic pentagon-only grain boundary (see Fig. \ref{fig:Scheme-3FPentOnly}). We start by writing the tight-binding 
equations in the grain boundary region\cite{Rodrigues_GBs-CA_PRB:2012,Rodrigues_GBs-TB_JPCM:2013,Paez_PRB:2015}
(we neglect out-of-plane relaxations in the GB region)
\begin{subequations} \label{eq:DfcTBeqs}
  \begin{eqnarray}
    -\frac{\epsilon}{t}\mathbf{A}(1) &=& W_{A}^{\dagger}\mathbf{B}(0)+\mathbf{B}(1) \,,\\
    -\frac{\epsilon}{t}\mathbf{B}(0) &=& \mathbf{D}(0)+W_{A}\mathbf{A}(1) \,,\\
    -\frac{\epsilon}{t}\mathbf{D}(0) &=& \mathbf{B}(0)+\sigma_{x}'\mathbf{D}(0)+\mathbf{A}(0),\\
    -\frac{\epsilon}{t}\mathbf{A}(0) &=& \mathbf{D}(0)+W_{A}^{\dagger}\mathbf{B}(-1) \,,\\
    -\frac{\epsilon}{t}\mathbf{B}(-1) &=& \mathbf{A}(-1)+W_{A}\mathbf{A}(0) \,,
  \end{eqnarray}
\end{subequations}
where the matrix $\sigma_{x}'$ is written in Eq. (\ref{eq:sigmaxp}). Note that $\sigma_{x}'$ depends both on
$k_{x} a$ and on the hopping parameters at the grain boundary, $\xi_{1}$, $\xi_{2}$ and $\xi_{3}$.

We can express these equations (see Appendix \ref{app:defectTB}) as a boundary condition equation connecting
the electronic wave function on the two sides of the grain boundary
\begin{eqnarray}
  \mathbf{L}(1) &=& \mathbb{M} \, \mathbf{L}(-1) \,. \label{eq:TBDC}
\end{eqnarray}
The boundary condition matrix $\mathbb{M}$ above is a $6\times6$ matrix given by
\begin{eqnarray}
  \mathbb{M} &=& R \mathbb{N}_{1} \mathbb{N}_{2} \mathbb{N}_{3} \mathbb{N}_{1} \mathbb{N}_{2} 
  R^{T} \,, \label{eq:TBDCMat}
\end{eqnarray}
where, for the sake of simplicity, we have omitted the dependence of the matrices $\mathbb{M}$ 
and $\mathbb{N}_{i}$ on $\epsilon/t$, $k_{x}$ and $\xi_{i}$. The matrices $\mathbb{N}_{i}$ are written in Eqs. 
(\ref{eq:NmatricesDfcLn}).

Finally, we can write the boundary condition matrix $\mathbb{M}$ in the basis uncoupling the three pairs of modes of 
the transfer matrix, $\widetilde{\mathbb{M}} = \Lambda(k_{x}a) \mathbb{M} [\Lambda(k_{x}a)]^{-1}$. By inspecting the
boundary condition Eq. (\ref{eq:TBDC}) in this basis, $\widetilde{\mathbf{L}}(1) = \widetilde{\mathbb{M}} \, 
\widetilde{\mathbf{L}}(-1)$, we readily conclude that, in general, $\widetilde{\mathbb{M}}$ mixes all the 
three modes of the transfer matrix (both the high-energy and the two low-energy ones).

\subsection{The boundary condition in the low-energy approximation}

\label{sec:CA-atGB}

At very low energies, $\epsilon \to 0$, and very near the Dirac points, $\mathbf{k} \to \mathbf{K}_{\nu} 
= (0, - 1) \nu 4 \pi / (3 \sqrt{3} a)$, the matrix $\widetilde{\mathbb{M}}$ acquires a somewhat simple
form -- see Eq. (\ref{eq:MatDiracPnt}). In this limit, the high-energy modes are evanescent, one of them
increasing and the other one decreasing exponentially with $n$ -- see Eq. (\ref{eq:TMdiags1-LowE-limit}) and 
subsequent paragraph. Since we must require the wave function to be normalizable, we conclude that when
$\epsilon \to 0$ and $\mathbf{k} \to \mathbf{K}_{\nu}$ the wave function must have the following form
\begin{subequations} \label{eq:LowEn-states}
  \begin{eqnarray}
    \widetilde{\mathbf{L}}(n) &\approx& \widetilde{\mathbf{l}}(n) = \left [ \begin{array}{c} 
    A_{h}(n) \\ 0 \\ A_{l-}(n) \\ B_{l-}(n) \\ A_{l+}(n) \\ B_{l+}(n) \end{array} 
    \right ] , \,\, \textrm{for $n>0$,}
  \end{eqnarray}
  and 
  \begin{eqnarray}
    \widetilde{\mathbf{L}}(n) &\approx& \widetilde{\mathbf{l}}(n) = \left [ \begin{array}{c} 
    0 \\ B_{h}(n) \\ A_{l-}(n) \\ B_{l-}(n) \\ A_{l+}(n) \\ B_{l+}(n) \end{array} 
    \right ] , \,\, \textrm{for $n<0$,}
  \end{eqnarray}
\end{subequations}
where, in order to keep the notation lighter, we have omitted the dependence on $k_{x}$ of both the vectors 
$\widetilde{\mathbf{L}}$ and $\widetilde{\mathbf{l}}$, and of the amplitudes $A$ and $B$.

We can now substitute Eqs. (\ref{eq:LowEn-states}) in the boundary condition $\widetilde{\mathbf{L}}(1) = 
\widetilde{\mathbb{M}} \, \widetilde{\mathbf{L}}(-1)$ to eliminate the high energy modes from our problem, 
ending up with an effective boundary condition that only involves the low-energy modes. Such a manipulation
generates an effective boundary condition matrix $\mathbb{M}^{\textrm{eff}}$ that we can express in terms of
the matrix elements of $\widetilde{\mathbb{M}}$ as follows
\begin{widetext}
\begin{eqnarray}
  \mathbb{M}^{\textrm{eff}} &=& \left[\begin{array}{cccc}
      \widetilde{\mathbb{M}}_{33}-\frac{\widetilde{\mathbb{M}}_{32} \widetilde{\mathbb{M}}_{23}}
      {\widetilde{\mathbb{M}}_{22}} & \widetilde{\mathbb{M}}_{34}-\frac{\widetilde{\mathbb{M}}_{32} 
        \widetilde{\mathbb{M}}_{24}}{\widetilde{\mathbb{M}}_{22}} & \widetilde{\mathbb{M}}_{35}-\frac{
        \widetilde{\mathbb{M}}_{32} \widetilde{\mathbb{M}}_{25}}{\widetilde{\mathbb{M}}_{22}} & 
      \widetilde{\mathbb{M}}_{36}-\frac{\widetilde{\mathbb{M}}_{32} \widetilde{\mathbb{M}}_{26}}
      {\widetilde{\mathbb{M}}_{22}} \\
      \widetilde{\mathbb{M}}_{43}-\frac{\widetilde{\mathbb{M}}_{42}\widetilde{\mathbb{M}}_{23}}
      {\widetilde{\mathbb{M}}_{22}} & \widetilde{\mathbb{M}}_{44}-\frac{\widetilde{\mathbb{M}}_{42}
        \widetilde{\mathbb{M}}_{24}}{\widetilde{\mathbb{M}}_{22}} & \widetilde{\mathbb{M}}_{45}-\frac{
        \widetilde{\mathbb{M}}_{42}\widetilde{\mathbb{M}}_{25}}{\widetilde{\mathbb{M}}_{22}} & 
      \widetilde{\mathbb{M}}_{46}-\frac{\widetilde{\mathbb{M}}_{42}\widetilde{\mathbb{M}}_{26}}{
        \widetilde{\mathbb{M}}_{22}} \\
      \widetilde{\mathbb{M}}_{53}-\frac{\widetilde{\mathbb{M}}_{52}\widetilde{\mathbb{M}}_{23}}{
        \widetilde{\mathbb{M}}_{22}} & \widetilde{\mathbb{M}}_{54}-\frac{\widetilde{\mathbb{M}}_{52}
        \widetilde{\mathbb{M}}_{24}}{\widetilde{\mathbb{M}}_{22}} & \widetilde{\mathbb{M}}_{55}-\frac{
        \widetilde{\mathbb{M}}_{52}\widetilde{\mathbb{M}}_{25}}{\widetilde{\mathbb{M}}_{22}} & 
      \widetilde{\mathbb{M}}_{56}-\frac{\widetilde{\mathbb{M}}_{52}\widetilde{\mathbb{M}}_{26}}{
        \widetilde{\mathbb{M}}_{22}} \\
      \widetilde{\mathbb{M}}_{63}-\frac{\widetilde{\mathbb{M}}_{62}\widetilde{\mathbb{M}}_{23}}{
        \widetilde{\mathbb{M}}_{22}} & \widetilde{\mathbb{M}}_{64}-\frac{\widetilde{\mathbb{M}}_{62}
        \widetilde{\mathbb{M}}_{24}}{\widetilde{\mathbb{M}}_{22}} & \widetilde{\mathbb{M}}_{65}-\frac{
        \widetilde{\mathbb{M}}_{62}\widetilde{\mathbb{M}}_{25}}{\widetilde{\mathbb{M}}_{22}} & 
      \widetilde{\mathbb{M}}_{66}-\frac{\widetilde{\mathbb{M}}_{62}\widetilde{\mathbb{M}}_{26}}{
        \widetilde{\mathbb{M}}_{22}}
    \end{array}\right] \,. \label{eq:MeffGen}
\end{eqnarray}
\end{widetext}
The matrix $\mathbb{M}^{\textrm{eff}}$ for the case of the 3-periodic pentagon-only grain boundary is written in
Eqs. (\ref{eq:Meff}) and (\ref{eq:fandg}).

As is widely known, in the low-energy continuum limit the tight-binding amplitudes, $C(\mathbf{r})$, can be expressed 
in terms of slowly varying fields, $\psi_{c}(\mathbf{r})$, as
\begin{eqnarray}
  C(\mathbf{r}) &\approx& \sum_{\nu=\pm1} e^{i \mathbf{K}_{\nu} \cdot \mathbf{r}} \psi_{c}^{\nu}(\mathbf{r}) \,. \label{eq:TB=D}
\end{eqnarray}
We can thus cast the tight-binding 4-spinor valid at low-energies, $\boldsymbol{\ell}(\mathbf{r}) = [A_{l-}(\mathbf{r}),
B_{l-}(\mathbf{r}),A_{l+}(\mathbf{r}),B_{l+}(\mathbf{r})]^{T}$, in terms of slowly varying Dirac fields
as $\boldsymbol{\ell}(\mathbf{r}) \approx [e^{i \mathbf{K}_{-} \cdot \mathbf{r}} \psi_{-}(\mathbf{r}), e^{i \mathbf{K}_{+} \cdot \mathbf{r}} 
\psi_{+}(\mathbf{r})]^{T}$, where $\psi_{\nu}(\mathbf{r}) = [\psi_{a_{\nu}},\psi_{b_{\nu}}]^{T}$.

We can finally write the boundary condition that Dirac fermions {\it see} at the 3-periodic pentagon-only grain boundary
as $\Psi(x,0^{+}) = \mathcal{M} \Psi(x,0^{-})$, where $\Psi(\mathbf{r}) = [\Psi_{-}(\mathbf{r}),\Psi_{+}(\mathbf{r})]^{T} =
[\psi_{a_{-}}(\mathbf{r}),\psi_{b_{-}}(\mathbf{r}),\psi_{a_{+}}(\mathbf{r}),\psi_{b_{+}}(\mathbf{r})]^{T}$. 

In the case of the 3-periodic pentagon-only grain boundary, the matrix $\mathcal{M}$ (see Appendix \ref{app:BC-CA}) reads
\begin{eqnarray}
  \mathcal{M} &=& \left [ \begin{array}{cccc} 0 & 1 & 0 & 0 \\ -1 & f & 0 & g^{*} \, e^{i \frac{2 \pi}{3}} 
       \\ 0 & 0 & 0 & 1 \\ 0 & g \, e^{-i \frac{2 \pi}{3}} & -1 & f \end{array} \right ] \,, \label{eq:BCmatrixP3}
\end{eqnarray}
where $f \equiv f(\xi_{1},\xi_{2},\xi_{3})$ and $g \equiv g(\xi_{1},\xi_{2},\xi_{3})$ are written in Eqs. (\ref{eq:fandg}).

Interestingly, in Eq. (\ref{eq:BCmatrixP3}) we clearly see that the off-diagonal blocks of matrix $\mathcal{M}$, those 
which control the intervalley scattering, are not {\it a priori} zero. We can thus conclude that in general this grain 
boundary gives rise to intervalley scattering. In particular, the intervalley scattering mixes the $\psi_{b}$ component 
of one valley with the same component of the other valley.

Notice that when all the three hoppings are equal (i.e. $\xi_{1}=\xi_{2}=\xi_{3}=\xi$), we get back to the simple case of 
the pentagon-only grain boundary (with periodicity $p=1$) that, as we know,\cite{Rodrigues_GBs-CA_PRB:2012} 
does not give rise to intervalley scattering. Owing to the fact that $f(\xi,\xi,\xi)=\xi$ and $g(\xi,\xi,\xi)=0$ [see Eqs. 
(\ref{eq:fandg})], its boundary condition matrix reads
\begin{eqnarray}
  \mathcal{M} &=& \left [ \begin{array}{cccc} 0 & 1 & 0 & 0 \\ -1 & \xi & 0 & 0 \\ 0 & 0 & 0 & 1 \\ 
      0 & 0 & -1 & \xi \end{array} \right ] \,. \label{eq:BCmatrix-SimplePO}
\end{eqnarray}
But this is natural since in such a case we are effectively dealing with a grain boundary with periodicity $\mathbf{u}_{1}$ 
which maps the projected Dirac points into distinct values of $k_{x}$ -- see Fig. \ref{fig:FBZ3-projected}. Similarly, when
we set $\xi_{1} = \xi_{2} = \xi$ and $\xi_{3}=0$, we recover the case of the $zz(558)$ grain boundary, which owing to its 
periodicity of $2 \mathbf{u}_{1}$ also maps the projected Dirac points into distinct values of $k_{x}$, which ends up blocking
intervalley scattering.\cite{Gunlycke_PRL:2011,Jiang_PLA:2011,Liwei_PRB:2012,Rodrigues_GBs-CA_PRB:2012,Rodrigues_GBs-TB_JPCM:2013,Ebert_JPCM:2014,Paez_PRB:2015}

Moreover, there are a few cases where, despite the 3-periodicity of the grain boundary,
intervalley scattering is suppressed. In these cases, the microscopic details of the grain boundary, i.e. the precise values
of $\xi_{1}$, $\xi_{2}$ and $\xi_{3}$, force $g(\xi_{1},\xi_{2},\xi_{3})=0$ thus forbidding intervalley scattering. Examples of
such cases are: $\xi_{3} = -3 \xi_{1}$ and $\xi_{2} = \xi_{1}$; $\xi_{3} = -\xi_{1}/3$ and $\xi_{2} = -\xi_{1}/3$; and 
$\xi_{3} = \xi_{1}$ and $\xi_{2} = -3 \xi_{1}$.

In the context of the perspective where we consider the grain boundary to be a finite width strip with a generalized potential 
[see Eq. (\ref{eq:GenPotV}) and end of Section \ref{sec:LowEn-3Periodic}], we show in Appendix \ref{app:GenPot} that the 
generalized potential originating from the 3-periodic pentagon-only grain boundary both has valley preserving terms (such as 
$V_{0 0}$ - a scalar potential - and $V_{0 1}$ - a constant gauge potential), and valley mixing terms (such as $V_{1 0}$, $V_{2 0}$,
$V_{1 1}$ and $V_{2 1}$). As a consequence, in general, a 3-periodic pentagon-only grain boundary will not only generate
intervalley scattering, but it will also prevent the existence of an angle of perfect transmission (see Appendix 
\ref{app:GenPot}).

\subsection{The Transmittance}

\label{sec:Transmittance}

As discussed in Section \ref{sec:LowEn-3Periodic} we can now compute the transmission and reflection coefficients 
$\tau_{\pm \pm}$, $\tau_{\pm \mp}$, $\rho_{\pm \pm}$ and $\rho_{\pm \mp}$. In particular, the transmission probability for an 
incoming electron living on the $\mathbf{K}_{+}$ ($\mathbf{K}_{-}$) valley to be transmitted into the same valley is given 
by $T_{++} = \vert \tau_{++} \vert^{2}$ ($T_{--} = \vert \tau_{--} \vert^{2}$), while the probability for it to be transmitted
into the other valley is given by $T_{-+} = \vert \tau_{-+} \vert^{2}$ ($T_{+-} = \vert \tau_{+-} \vert^{2}$).

\begin{figure}
  \includegraphics[width=0.95\columnwidth]{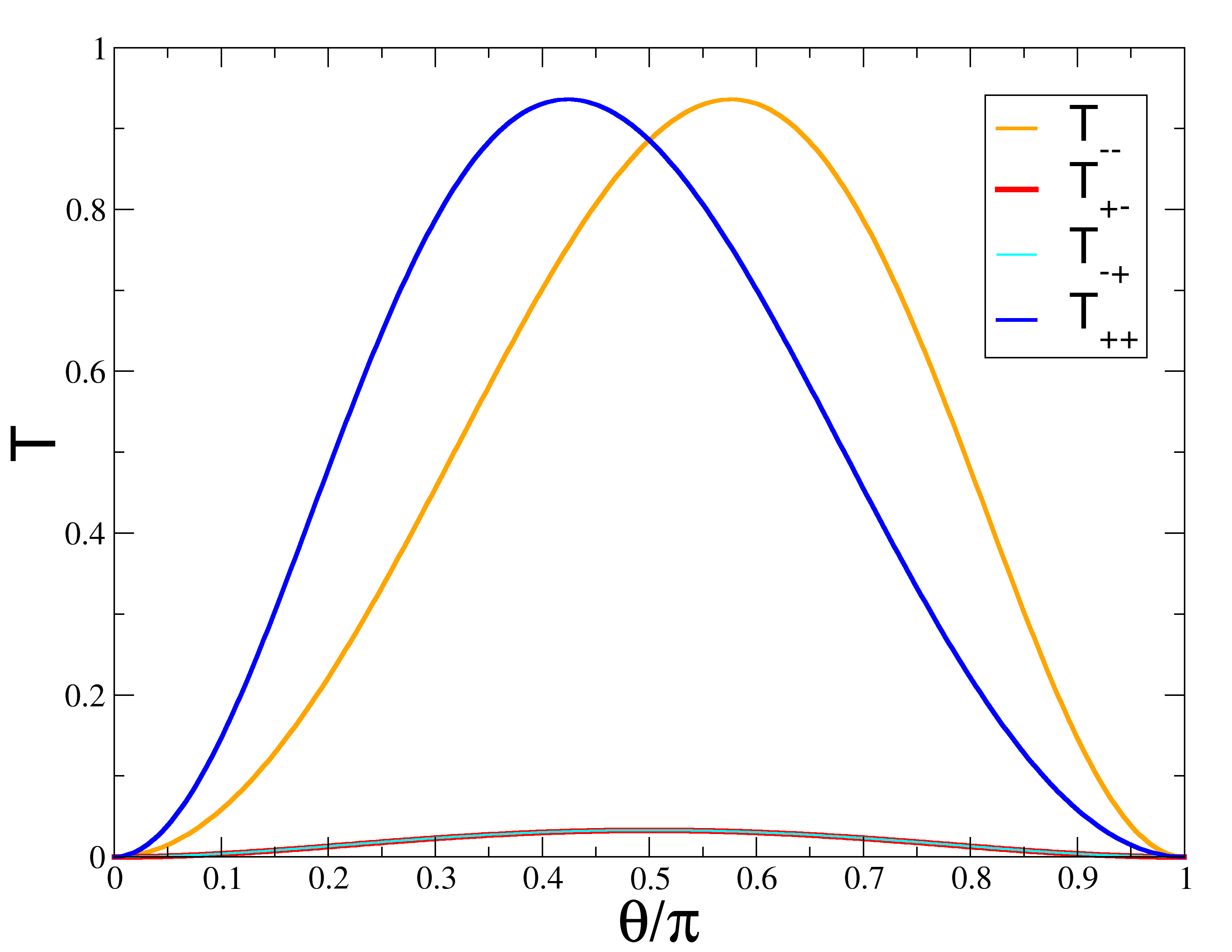}
  \caption{Massless Dirac fermions' transmission probabilities for the
    3-periodic pentagon-only grain boundary in terms of the angle of
    incidence, $\theta$. The $T_{--}$ ($T_{+-}$) [$T_{-+}$ ($T_{++}$)]
    stands for the probability of transmission of an incident Dirac
    fermion living on the $\mathbf{K}_{-}$ [$\mathbf{K}_{+}$] valley
    to be transmitted into the $\mathbf{K}_{-}$ ($\mathbf{K}_{+}$)
    valley. The hopping parameters at the grain boundary were set to:
    $\xi_{1}=0.1 t$, $\xi_{2}=0.4 t$ and $\xi_{3}=0.8 t$.}
  \label{fig:Transmittances-3FPentOnly}
\end{figure}

In Eqs. (\ref{eq:transmission-Coefs}) we write the expressions of the $\tau_{++}$, $\tau_{-+}$, $\tau_{+-}$ and $\tau_{--}$, 
for the 3-periodic pentagon-only grain boundary with general hoppings (in units of $t$), $\xi_{1}$, $\xi_{2}$ and $\xi_{3}$. 
These were obtained by solving the system of linear equations defined by $\Psi(x,0^{+}) = \mathcal{M} \Psi(x,0^{-})$ where 
$\mathcal{M}$ is given by Eq. (\ref{eq:BCmatrixP3}). In Fig. \ref{fig:Transmittances-3FPentOnly} we have plotted the 
transmission probabilities $T_{\pm \pm}$ and $T_{\pm \mp}$ for the case where the hopping parameters are $\xi_{1}=0.1$, 
$\xi_{2}=0.4$ and $\xi_{3}=0.8$.

\begin{figure}
  \includegraphics[width=0.95\columnwidth]{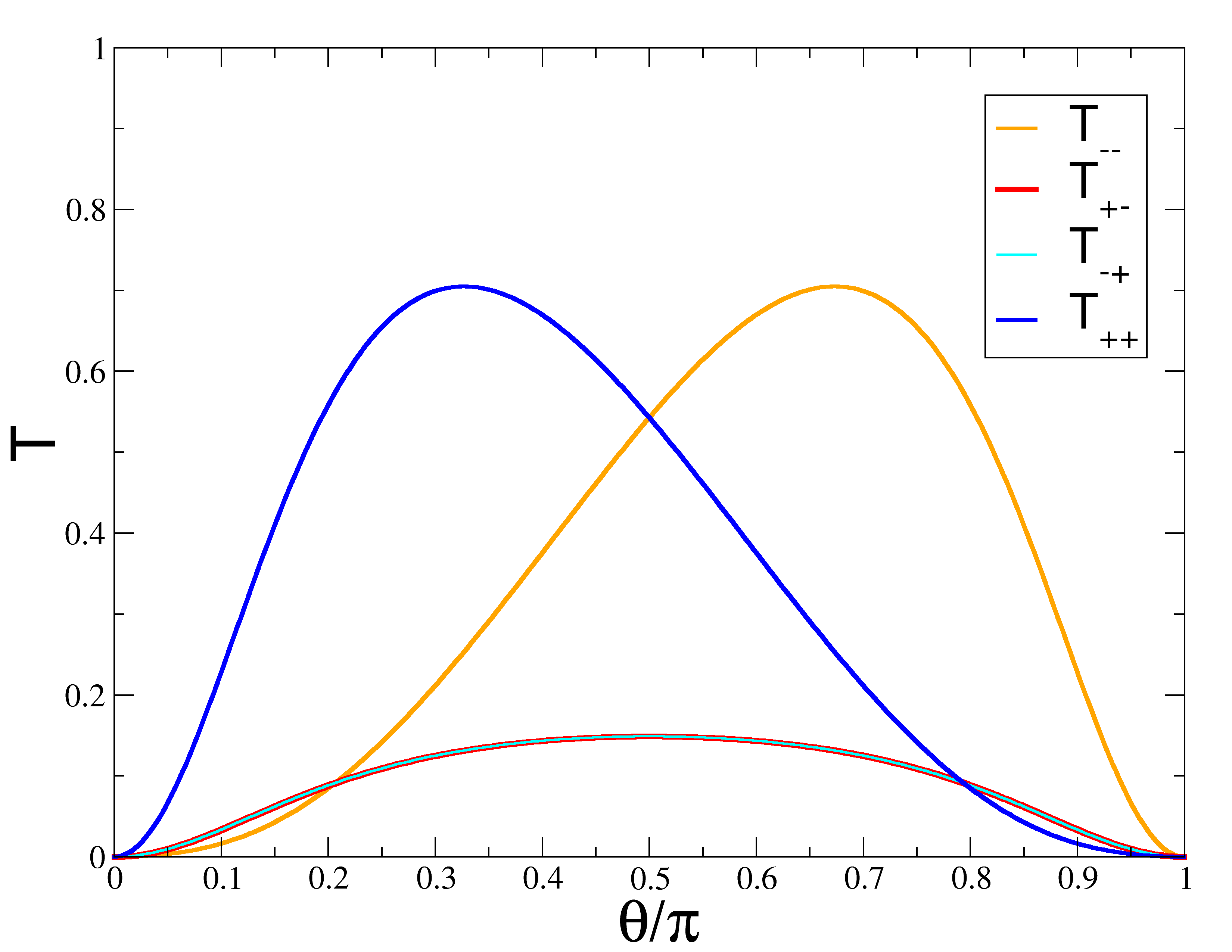}
  \caption{Same as in Fig. \ref{fig:Transmittances-3FPentOnly}, but
    now with the hopping parameters at the grain boundary set to:
    $\xi_{1}=1.1 t$, $\xi_{2}=0.05 t$ and $\xi_{3}=2.2 t$.}
  \label{fig:Transmittances-3FPentOnly__2}
\end{figure}

For a 3-periodic pentagon-only grain boundary with the above hopping parameters, the intervalley scattering is
weak, with the intervalley transmission probabilities being considerably smaller than the valley preserving ones. 
However, this picture can be greatly modified if we choose an appropriate set of hopping parameters at the 
grain boundary. As an example, in Fig. \ref{fig:Transmittances-3FPentOnly__2} we plot the transmission 
probabilities for a case where $\xi_{1}=1.1$, $\xi_{2}=0.05$ and $\xi_{3}=2.2$, which shows much stronger intervalley 
scattering than Fig. \ref{fig:Transmittances-3FPentOnly}.

This robust increase of the intervalley transmission (compare Figs. \ref{fig:Transmittances-3FPentOnly} and
\ref{fig:Transmittances-3FPentOnly__2}) can be traced back to a strong amplification  of the terms $V_{1 0}$ and 
$V_{1 1}$ of the generalized potential existent inside the strip that mimics the effect of the grain boundary 
in the low-energy limit. Such an amplification (by nearly two orders of magnitude) gives rise to a set of 
generalized potential terms of similar magnitudes, thus increasing the amount of valley mixing of the 
eigenmodes living inside the strip. The stronger the valley mixing of these modes, the more the electron's valley 
quantum number rotates while propagating inside the strip, and thus more wave function weight is transferred
between valleys while the Dirac fermion propagates inside the strip.

Finally, we must note that, as we can quickly infer from the above results (where we have considered different 
sets of values for the hopping renormalizations $\xi_{i}$), the scattering properties of the grain boundary
are strongly dependent on the microscopic details at the grain boundary, as previously remarked by other
studies.\cite{Rodrigues_GBs-CA_PRB:2012,Rodrigues_GBs-TB_JPCM:2013,Paez_PRB:2015} Such behaviour points towards
the possibility of making use of this kind of nanostructures to control and explore the valley degree of freedom 
of graphene. Chemical decoration of the grain boundary region, application of strains, of electric and of magnetic 
fields, all likely modify the electronic scattering off these grain boundaries, thus suggesting their usage as 
sensors and current switchers.

\section{Conclusion}

\label{sec:conclusion}

To close, let us briefly summarize the contents of this manuscript. We have started by analyzing in general terms the 
low-energy charge carrier transport across zigzag-aligned degenerate 3-periodic grain boundaries. We have then 
demonstrated such results by working out the low-energy charge transport across a 3-periodic pentagon-only grain boundary. 
In particular, starting from its microscopic tight-binding model, we have derived the boundary condition seen by the 
massless Dirac fermions at such grain boundary. With it we have calculated the scattering coefficients, from which we 
concluded that the valley preserving and intervalley scattering probabilities are highly responsive to external manipulation
through control of the grain boundary's microscopic details. We have also made use of the generalized potential representation 
of the grain boundary to gain insight into the obtained results.

\begin{acknowledgements} 
      J. N. B. R. thanks Jo\~ao M. B. Lopes dos Santos and Nuno M. R. Peres for many insightful discussions and helpful 
      comments. The author also acknowledges the Faculdade de Ci\^encias da Universidade do Porto for the warm support 
      during the initial stages of this work, as well as the Singapore National Research Foundation, the Prime Minister 
      Office, the Singapore Ministry of Education and Yale-NUS College for their Fellowship Programs (NRF-NRFF2012-01, 
      R-144-000-295-281 and R-607-265-01312).     
\end{acknowledgements}

\appendix

\section{Tight-binding model}

\label{app:TBtechs}

In this section we will give details of the derivations presented in Section \ref{sec:TBmodel-3PO} concerning 
the microscopic tight-binding model of monolayer graphene with a 3-periodic pentagon-only grain boundary. The calculations
below closely follow what was done in Ref. \onlinecite{Paez_PRB:2015}. In sub-Section \ref{app:BulkTB} we 
concentrate on the calculations leading to the bulk transfer matrix, while in sub-Section \ref{app:defectTB}
we focus on the calculations giving rise to the tight-binding boundary condition that originates from the presence 
of the 3-periodic pentagon-only grain boundary.

\subsection{The tight-binding equations at the bulk}

\label{app:BulkTB}

The matrix $W_{A}$ present in the bulk tight-binding equations [see Eqs. (\ref{eq:BulkTBeqs})] reads
\begin{eqnarray}
  W_{A} &=& \left[\begin{array}{ccc}
      1 & 1 & 0 \\
      0 & 1 & 1 \\
      e^{3 i k_{x}a} & 0 & 1
    \end{array}\right] \,. \label{eq:WAmatrix}
\end{eqnarray}

The bulk tight-binding equations [see Eqs. (\ref{eq:BulkTBeqs})] can be cast in the form
\begin{subequations} \label{eq:DfcTBeqsBulk}
  \begin{eqnarray}
    \left[\begin{array}{c}
        \mathbf{B}(n)\\
        \mathbf{A}(n)
      \end{array}\right] &=& \mathbb{Q}_{1} \left[\begin{array}{c}
        \mathbf{A}(n)\\
        \mathbf{B}(n-1)
      \end{array}\right] \,,\label{eq:Bulk1}\\
    \left[\begin{array}{c}
        \mathbf{A}(n)\\
        \mathbf{B}(n-1)
      \end{array}\right] &=& \mathbb{Q}_{2} \left[\begin{array}{c}
        \mathbf{B}(n-1)\\
        \mathbf{A}(n-1)
      \end{array}\right] \,, \label{eq:Bulk2}
  \end{eqnarray}
\end{subequations}
where the matrices $\mathbb{Q}_{1}$ and $\mathbb{Q}_{2}$ read
\begin{subequations} \label{eq:Q1Q2}
  \begin{eqnarray}
    \mathbb{Q}_{1} &=& -\left[\begin{array}{cc}
        \frac{\epsilon}{t}\mathbb{I}_{3} & W_{A}^{\dagger}\\
        -\mathbb{I}_{3} & 0
      \end{array}\right] \,, \\
    \mathbb{Q}_{2} &=& -\left[\begin{array}{cc}
        \frac{\epsilon}{t}\big(W_{A}\big)^{-1} & \big(W_{A}\big)^{-1}\\
        -\mathbb{I}_{3} & 0
      \end{array}\right] \,,
  \end{eqnarray}
\end{subequations}
with $\mathbb{I}_{3}$ standing for the $3 \times 3$ unit matrix.

Eqs. (\ref{eq:DfcTBeqsBulk}) can be written in the form of a transfer matrix, as was done in Eq. (\ref{eq:transfMatEq}),
where the transfer matrix is given by Eq. (\ref{eq:TransferMatrixCoup}). The matrix $R$ present in the latter equation
reads
\begin{eqnarray}
  R &=& \left[\begin{array}{cccccc}
      0 & 0 & 0 & 1 & 0 & 0 \\
      1 & 0 & 0 & 0 & 0 & 0 \\
      0 & 0 & 0 & 0 & 1 & 0 \\
      0 & 1 & 0 & 0 & 0 & 0 \\
      0 & 0 & 0 & 0 & 0 & 1 \\
      0 & 0 & 1 & 0 & 0 & 0
    \end{array}\right] \,. \label{eq:Rmatrix}
\end{eqnarray}
This matrix simply changes from the basis $\{B_{1}(n),B_{2}(n),B_{3}(n),A_{1}(n),A_{2}(n),A_{3}(n) \}$ into the basis
$\{ A_{1}(n),B_{1}(n),A_{2}(n),B_{2}(n),A_{3}(n),B_{3}(n) \}$.

The matrix $\Lambda(\phi)$ enforcing the basis change that uncouples the modes of the transfer matrix reads
\begin{widetext}
  \begin{eqnarray}
    \Lambda(\phi) &=& \frac{1}{\sqrt{3}} \left [ \begin{array}{cccccc} 1 & 0 & - \frac{e^{-i (\phi - 2 \pi/3)} 
          i \sqrt{3}}{1 + e^{i \pi/3}} & 0 & \frac{e^{-i 2 (\phi - 2 \pi/3)} i \sqrt{3}}{1 + e^{-i \pi/3}} & 0 \\ 
        0 & 1 & 0 & - \frac{e^{-i (\phi - 2 \pi/3)} i \sqrt{3}}{1 + e^{i \pi/3}} & 0 & \frac{e^{-i 2 (\phi - 2 \pi/3)} 
          i \sqrt{3}}{1 + e^{-i \pi/3}} \\
        1 & 0 & -e^{-i (\phi - \pi/3)} & 0 & -e^{-i (2 \phi + \pi/3)} & 0 \\
        0 & 1 & 0 & -e^{-i (\phi - \pi/3)} & 0 & -e^{-i (2 \phi + \pi/3)} \\
        1 & 0 & -e^{-i (\phi + \pi/3)} & 0 & -e^{-i (2 \phi - \pi/3)} & 0 \\
        0 & 1 & 0 & -e^{-i (\phi + \pi/3)} & 0 & -e^{-i (2 \phi - \pi/3)} 
      \end{array} \right ] \,, \label{eq:LambdaMat}
  \end{eqnarray}
\end{widetext}
where $\phi = k_{x} a$.

In this basis, the three $2 \times 2$ matrices in the diagonal of the transfer matrix $\widetilde{\mathbb{T}}(\epsilon,k_{x}a)$
are noted by $\mathbb{T}_{h}$, $\mathbb{T}_{l-}$ and $\mathbb{T}_{l+}$. The matrix
\begin{subequations} \label{eq:TMdiags}
  \begin{eqnarray}
    \mathbb{T}_{h}(\epsilon,k_{x}a) &=& \frac{1}{1 + e^{i k_{x} a}} \left [ \begin{array}{cc} -1 & - \epsilon 
        \\ \epsilon & -2 - 2 \cos(k_{x} a) + \epsilon^{2} \end{array} \right ] \,, \nonumber \\ \label{eq:TMdiags1}
  \end{eqnarray}
  corresponds to the high-energy mode when we are around $k_{x} a = 0$. Similarly, the matrices corresponding to the 
  low-energy modes (one from each Dirac point) read
  \begin{eqnarray}
    \mathbb{T}_{l-}(\epsilon,k_{x}a) &=& \Upsilon(k_{x}) \left [ \begin{array}{cc} - 1 & - \epsilon \\ \epsilon 
        & \frac{e^{-i (k_{x} a - \frac{\pi}{3})}- 1}{f(k_{x})} + \epsilon^{2} \end{array} \right ] \,, \label{eq:TMdiags2} \\
    \mathbb{T}_{l+}(\epsilon,k_{x}a) &=& \Theta(k_{x}) \left [ \begin{array}{cc} -1 & -\epsilon \\ \epsilon & 
        \frac{e^{-i (k_{x} a + \frac{\pi}{3})}-1}{g(k_{x})} + \epsilon^{2} \end{array} \right ] \,, \label{eq:TMdiags3}
  \end{eqnarray}
\end{subequations}
where we have defined $\Upsilon(k_{x})$ and $\Theta(k_{x})$ as
\begin{subequations}
  \begin{eqnarray}
    \Upsilon(k_{x}) &=& \frac{e^{i \pi / 3} - e^{-i k_{x}a}}{1 - 2 \cos(k_{x} a)} \,, \\
    \Theta(k_{x}) &=& \frac{e^{-i \pi / 3} - e^{-i k_{x}a}}{1 - 2 \cos(k_{x} a)} \,.
  \end{eqnarray}
\end{subequations}

\subsection{The tight-binding equations at the grain boundary}

\label{app:defectTB}

The matrix $\sigma_{x}'$ present in the tight-binding equations at the grain boundary region [see Eqs. (\ref{eq:DfcTBeqs})]
reads
\begin{eqnarray}
  \sigma_{x}' &=& \left[\begin{array}{ccc}
      0 & \xi_{1} & \xi_{3} e^{-3 i k_{x}a} \\
      \xi_{1} & 0 & \xi_{2} \\
      \xi_{3} e^{3 i k_{x}a} & \xi_{2} & 0
    \end{array}\right] \,, \label{eq:sigmaxp}
\end{eqnarray}
where $\xi_{1}$, $\xi_{2}$ and $\xi_{3}$ stand for the hopping parameters at the grain boundary region as represented in
Fig. \ref{fig:Scheme-3FPentOnly}.  

Eqs. (\ref{eq:DfcTBeqs}) can be condensed in the form
\begin{subequations} \label{eq:DfcTBeqs3}
  \begin{eqnarray}
    \left[\begin{array}{c}
        \mathbf{B}(1)\\
        \mathbf{A}(1)
      \end{array}\right] &=& \mathbb{N}_{1} \left[\begin{array}{c}
        \mathbf{A}(1)\\
        \mathbf{B}(0)
      \end{array}\right] \,,\label{eq:BC1}\\
    \left[\begin{array}{c}
        \mathbf{A}(1)\\
        \mathbf{B}(0)
      \end{array}\right] &=& \mathbb{N}_{2} \left[\begin{array}{c}
        \mathbf{B}(0)\\
        \mathbf{D}(0)
      \end{array}\right],\label{eq:BC2}\\
    \left[\begin{array}{c}
        \mathbf{B}(0)\\
        \mathbf{D}(0)
      \end{array}\right] &=& \mathbb{N}_{3} \left[\begin{array}{c}
        \mathbf{D}(0)\\
        \mathbf{A}(0)
      \end{array}\right] \,,\label{eq:BC3}\\
    \left[\begin{array}{c}
        \mathbf{D}(0)\\
        \mathbf{A}(0)
      \end{array}\right] &=& \mathbb{N}_{1} \left[\begin{array}{c}
        \mathbf{A}(0)\\
        \mathbf{B}(-1)
      \end{array}\right] \,,\label{eq:BC4}\\
    \left[\begin{array}{c}
        \mathbf{A}(0)\\
        \mathbf{B}(-1)
      \end{array}\right] &=& \mathbb{N}_{2} \left[\begin{array}{c}
        \mathbf{B}(-1)\\
        \mathbf{A}(-1)
      \end{array}\right] \,.\label{eq:BC5}
  \end{eqnarray}
\end{subequations}
where the matrices $\mathbb{N}_{1}$, $\mathbb{N}_{2}$ and
$\mathbb{N}_{3}$ are $6 \times 6$ matrices which read
\begin{subequations} \label{eq:NmatricesDfcLn}
  \begin{eqnarray}
    \mathbb{N}_{1} &=& -\left[\begin{array}{cc}
        \frac{\epsilon}{t}\mathbb{I}_{3} & \big(W_{A}\big)^{\dagger}\\
        -\mathbb{I}_{3} & 0
      \end{array}\right] \,, \\
    \mathbb{N}_{2} &=& -\left[\begin{array}{cc}
        \frac{\epsilon}{t}\big(W_{A}\big)^{-1} & \big(W_{A}\big)^{-1}\\
        -\mathbb{I}_{3} & 0
      \end{array}\right] \,, \\
    \mathbb{N}_{3} &=& -\left[\begin{array}{cc}
        \Big(\frac{\epsilon}{t}\mathbb{I}_{3}+\sigma_{x}'\Big) & \mathbb{I}_{3}\\
        -\mathbb{I}_{3} & 0
      \end{array}\right] \,.
  \end{eqnarray}
\end{subequations}
The above matrices depend on the reduced energy, $\epsilon/t$, on the longitudinal momentum, $k_{x}$, and on the hopping
parameters at the defect, $\xi_{1}$, $\xi_{2}$ and $\xi_{3}$.

\section{The boundary condition matrix in the continuum approximation}

\label{app:BC-CA}

When $\epsilon \to 0$ and $\mathbf{k} \to \mathbf{K}_{\nu} = (0, - 1) \nu 4 \pi / (3 \sqrt{3} a)$, the boundary 
condition matrix (expressed in the basis uncoupling the modes of the transfer matrix $\mathbb{T}$) 
$\widetilde{\mathbb{M}}$ reads 
\begin{widetext}
\begin{eqnarray}
  \widetilde{\mathbb{M}} &=& \left[\begin{array}{cccccc}
      0 & 1 & 0 & 0 & 0 & 0 \\
      -1 & -\frac{8}{3} (\xi_{1}+\xi_{2}+\xi_{3}) & 0 & \frac{2}{3} (-\xi_{1}+\xi_{2} e^{-i \pi/3} + \xi_{3} e^{i \pi/3}) 
      & 0 & \frac{2}{3} (- \xi_{1}+\xi_{2} e^{i \pi/3} + \xi_{3} e^{-i \pi/3}) \\
      0 & 0 & 0 & -e^{i \pi/3} & 0 & 0 \\
      0 & \frac{2}{3} (\xi_{1} e^{i \pi / 3} + \xi_{2} e^{-i \pi/3} - \xi_{3}) & e^{i \pi/3} & -\frac{e^{i \pi/3}}{3} 
      (\xi_{1}+\xi_{2}+\xi_{3}) & 0 & \frac{2}{3} (\xi_{1} e^{i \pi / 3} - \xi_{2} + \xi_{3} e^{-i \pi/3}) \\
      0 & 0 & 0 & 0 & 0 & -e^{-i \pi /3} \\
      0 & \frac{2}{3} (\xi_{1} e^{-i \pi/3} + \xi_{2} e^{i \pi /3} - \xi_{3}) & 0 & \frac{2}{3} (\xi_{1} e^{-i \pi/3} 
      - \xi_{2} + \xi_{3} e^{i \pi/3}) & e^{-i \pi /3} & - \frac{e^{-i \pi/3}}{3} (\xi_{1}+\xi_{2}+\xi_{3})
    \end{array}\right] \,. \nonumber \\ \label{eq:MatDiracPnt}
\end{eqnarray}
\end{widetext}

In this limit, the $2 \times 2$ matrix $\mathbb{T}_{h}$ describing the high energy modes of the transfer matrix
$\widetilde{\mathbb{T}}$ [see Eq. (\ref{eq:TMdiags1})] reads
\begin{eqnarray}
  \mathbb{T}_{h}(0,0) &=& - \left [ \begin{array}{cc} \frac{1}{2} & 0 \\ 
  0 & 2 \end{array} \right ] \,. \nonumber \\ \label{eq:TMdiags1-LowE-limit}
\end{eqnarray}
It is then straightforward to understand what will be the relation between the high-energy modes' amplitudes
$\widetilde{\mathbf{L}}(i)$ at position $i$, and the amplitudes $\widetilde{\mathbf{L}}(j) = \widetilde{\mathbb{T}}^{n}
\widetilde{\mathbf{L}}(i)$ at position $j=i+n$: since $1/ 2 < 1$, the {\it upper} high-energy mode, i.e. $\psi_{h_{u}} = 
\{1, 0\}^{T}$, is going to decrease exponentially with $n$, while the {\it lower} one, i.e. $\psi_{h_{l}} = \{0, 1\}^{T}$, 
is going to exponentially increase because $2 > 1$.

Therefore, the requirement that the wave function be normalizable, implies that $\widetilde{L}(n)$ must have the form of Eq. 
(\ref{eq:LowEn-states}). Hence, as described in the main text, we can then eliminate the high-energy modes from the 
problem, and write the effective boundary condition seen by a low-energy electron (hole) inciding in the 3-periodic 
pentagon-only from infinity. In particular, the matrix $\mathbb{M}^{\textrm{eff}}(\epsilon=0,k_{x}a=0)$  obtained from 
Eq. (\ref{eq:MeffGen}) reads
  \begin{eqnarray}
    \mathbb{M}^{\textrm{eff}} &=& \left[\begin{array}{cccc}
        0 & e^{-i \frac{2 \pi}{3}} & 0 & 0 \\
        -e^{-i \frac{2 \pi}{3}} & e^{-i \frac{2 \pi}{3}} \, f & 0 & g^{*} \\
        0 & 0 & 0 & e^{i \frac{2 \pi}{3}} \\
        0 & g & - e^{i \frac{2 \pi}{3}} & e^{i \frac{2 \pi}{3}} \, f
      \end{array} \right] \,, \label{eq:Meff}
  \end{eqnarray}
where $f \equiv f(\xi_{1},\xi_{2},\xi_{3})$ and $g \equiv g(\xi_{1},\xi_{2},\xi_{3})$ can be written as
\begin{widetext}
\begin{subequations} \label{eq:fandg}
  \begin{eqnarray}
    f(\xi_{1},\xi_{2},\xi_{3}) &=& \frac{\xi_{1}^{2} + \xi_{2}^{2} + \xi_{3}^{2} + (\xi_{1} \xi_{2} 
      + \xi_{1} \xi_{3} + \xi_{2} \xi_{3})}{2 (\xi_{1} + \xi_{2} + \xi_{3})} \,, \\
    g(\xi_{1},\xi_{2},\xi_{3}) &=& \frac{e^{-i \pi / 3} \xi_{1}^{2} - \xi_{2}^{2} + e^{i \pi / 3} \xi_{3}^{2} 
      + 2 (e^{-i 2 \pi / 3} \xi_{1} \xi_{2} + \xi_{1} \xi_{3} + e^{i 2 \pi / 3} \xi_{2} \xi_{3})}{2 (\xi_{1} 
      + \xi_{2} + \xi_{3})} \,.
  \end{eqnarray}
\end{subequations}
\end{widetext}

The boundary condition seen by the massless Dirac fermions is finally given by substituting Eq. (\ref{eq:TB=D}) in 
$\boldsymbol{\ell}(m \mathbf{u}_{1} + \mathbf{u}_{2}) = \mathbb{M}^{\textrm{eff}} \, \boldsymbol{\ell}(m \mathbf{u}_{1} - 
\mathbf{u}_{2})$. Since $\mathbf{K}_{\nu} \cdot \mathbf{u}_{2} = - \nu 2 \pi / 3$, such condition can be recast
as
\begin{eqnarray}
    \Psi(x,0^{+}) &=& \Xi \, \mathbb{M}^{\textrm{eff}}(0,0) \, \Xi \, \Psi(x,0^{-}) \,, \label{eq:bc-provis}
\end{eqnarray}
where $\Psi(x,y) = [\psi_{a_{-}}, \psi_{b_{-}}, \psi_{a_{+}}, \psi_{b_{+}}]^{T}$ (we have omitted the dependence of the 
$\psi$ components on $x$ and $y$) and $\Xi \equiv \exp[i \frac{2 \pi}{3} \tau_{3} \otimes \sigma_{0}]$. Eq. 
(\ref{eq:bc-provis}) can be written as $\Psi(x,0^{+}) = \mathcal{M} \Psi(x,0^{-})$ with $\mathcal{M}$ given in Eq. 
(\ref{eq:BCmatrixP3}).

Finally, the transmission coefficients $\tau_{++}$, $\tau_{+-}$, $\tau_{-+}$ and $\tau_{--}$ for the 3-periodic pentagon-only 
grain boundary can be shown to have the following analytic expressions:
\begin{widetext}
  \begin{subequations} \label{eq:transmission-Coefs}
    \begin{eqnarray}
      \tau_{--} &=& - e^{-i \theta} \sin \theta \frac{\xi_{1}^{2} + \xi_{2}^{2} + \xi_{3}^{2} 
        + \xi_{1} \xi_{2} + \xi_{1} \xi_{3} + \xi_{2} \xi_{3} - 4 e^{i \theta} (\xi_{1} + \xi_{2} + \xi_{3})}{- i 4 
        (\xi_{1} + \xi_{2} + \xi_{3}) + i 3 \xi_{1} \xi_{2} \xi_{3} + 2 \sin \theta \big(\xi_{1}^{2} + \xi_{2}^{2} 
        + \xi_{3}^{2} + (\xi_{1} \xi_{2} + \xi_{1} \xi_{3} + \xi_{2} \xi_{3}) \big)} \,, \\
      \tau_{-+} &=& e^{-i \theta} \sin \theta \frac{- e^{i 2 \pi / 3} \xi_{1}^{2} - \xi_{2}^{2} - e^{-i 2 \pi / 3} 
        \xi_{3}^{2} + 2 ( e^{-i 2 \pi / 3} \xi_{1} \xi_{2} + \xi_{1} \xi_{3} + e^{i 2 \pi / 3} \xi_{2} \xi_{3})}{- i 4 
        (\xi_{1} + \xi_{2} + \xi_{3}) + i 3 \xi_{1} \xi_{2} \xi_{3} + 2 \sin \theta \big(\xi_{1}^{2} + \xi_{2}^{2} 
        + \xi_{3}^{2} + (\xi_{1} \xi_{2} + \xi_{1} \xi_{3} + \xi_{2} \xi_{3}) \big)} \,, \\
      \tau_{+-} &=& e^{i \theta} \sin \theta \frac{e^{-i 2 \pi/3} \xi_{1}^{2} + \xi_{2}^{2} + e^{i 2 \pi/3} 
        \xi_{3}^{2} - 2 ( e^{i 2 \pi/3} \xi_{1} \xi_{2} + \xi_{1} \xi_{3} + e^{-i 2 \pi/3} \xi_{2} \xi_{3})}{- i 4 
        (\xi_{1} + \xi_{2} + \xi_{3}) + i 3 \xi_{1} \xi_{2} \xi_{3} + 2 \sin \theta \big(\xi_{1}^{2} + \xi_{2}^{2} 
        + \xi_{3}^{2} + (\xi_{1} \xi_{2} + \xi_{1} \xi_{3} + \xi_{2} \xi_{3}) \big)} \,, \\
      \tau_{++} &=& e^{i \theta} \sin \theta \frac{\xi_{1}^{2} + \xi_{2}^{2} + \xi_{3}^{2} + \xi_{1} \xi_{2} 
        + \xi_{1} \xi_{3} + \xi_{2} \xi_{3} + 4 e^{-i \theta} ( \xi_{1} + \xi_{2} + \xi_{3})}{- i 4 (\xi_{1} 
        + \xi_{2} + \xi_{3}) + i 3 \xi_{1} \xi_{2} \xi_{3} + 2 \sin \theta \big(\xi_{1}^{2} + \xi_{2}^{2} + \xi_{3}^{2} 
        + (\xi_{1} \xi_{2} + \xi_{1} \xi_{3} + \xi_{2} \xi_{3}) \big)} \,.
    \end{eqnarray}
  \end{subequations}
\end{widetext}

It is possible to show that, for the 3-periodic pentagon-only grain boundary with any hopping parameters values $\xi_{1}$,
$\xi_{2}$ and $\xi_{3}$, the intervalley scattering is always the same for incoming electrons either living on the valley 
$\mathbf{K}_{-}$ or on the valley $\mathbf{K}_{+}$, i.e. $T_{+-} = \vert \tau_{+-} \vert^{2} = \vert \tau_{-+} \vert^{2} 
= T_{-+}$.

Similarly, we can also show that $\tau_{+-}(\theta) = \tau_{-+}(\pi-\theta)$, and thus the transmission plots (see Figs. 
\ref{fig:Transmittances-3FPentOnly} and \ref{fig:Transmittances-3FPentOnly__2}) are always symmetric (upon interchanging 
of valley) over the $\theta=\pi/2$ angle.

\section{The boundary condition matrix in terms of the generalized potential}

\label{app:GenPot}

In this appendix we will show how can we connect the two perspectives discussed in Section \ref{sec:LowEn-3Periodic}
for the low-energy electronic scattering off a periodic grain boundary. In particular, we will show how can we compute
the generalized potential $\mathbb{V}$ in Eq. (\ref{eq:GenPotV}) in terms of the boundary condition matrix $\mathcal{M}$ 
originating from the tight-binding model of the grain boundary.

As discussed in Section \ref{sec:LowEn-3Periodic}, in the low energy continuum limit of the tight-binding model we 
can see the grain boundary as a finite width strip where the Dirac fermions are governed by the following Hamiltonian
\begin{eqnarray}
  H &=& v_{F} ( \tau_{3} \otimes \sigma_{1}, \tau_{0} \otimes \sigma_{2} ) \cdot \mathbf{p} + \mathbb{V} \,, \label{eq:GenHam}
\end{eqnarray} 
where the $\tau_{i}$ and $\sigma_{i}$ ($i=1,2,3$) stand for the $2 \times 2$ Pauli matrices acting on, respectively, the valley 
and the pseudo-spin degrees of freedom. Similarly, $\tau_{0}$ and $\sigma_{0}$ stand for the $2 \times 2$ identity matrix acting
on each of these sub-spaces. Note that in the above equation (and in the remaining of this appendix) we have set $\hbar = 1$.

The term $\mathbb{V}$ in Eq. (\ref{eq:GenHam}) stands for a generalized potential acting on graphene's massless Dirac fermions. 
By forcing this generalized
potential $\mathbb{V}$ to be hermitian [see general expression in Eq. (\ref{eq:GenPotV})] we ensure that the boundary condition
matrix $\mathcal{M}$ conserves the flux, $\mathcal{M}^{\dagger} J_{y} \mathcal{M} = J_{y}$, as required. Furthermore, the time-reversal
invariance of $\mathcal{M}$ (whenever the grain boundary is non-magnetic) is ensured by requiring that $\mathbb{V}$ is also 
time-reversal invariant. 

Given this, and before proceeding, let us briefly analyze the effect of each of the terms $V_{\alpha \beta}$ on the eigenmodes 
living inside the finite width strip. The terms $V_{0 \beta}$ act equally on both valleys. The term proportional to $V_{0 0}$ 
represents an electrostatic potential analogous to that generated by gating graphene or by the presence of charge impurities 
in the vicinity of the graphene flake. The term $V_{0 3}$ is a mass 
term equivalent to that originating whenever the atoms of each sub-lattice have different onsite energies. Terms proportional 
to $V_{0 1}$ and $V_{0 2}$ are analogous to the $x$- and $y$-component of a vector potential $A_{x}$ and $A_{y}$ arising from 
the presence of a magnetic field perpendicular to the graphene layer. The terms $V_{3 \beta}$ can be viewed as analogous to
those originating from a pseudo-magnetic field generated by deformations of the honeycomb lattice. All the other terms, 
$V_{i \beta}$ (with $i=1, 2$ and $\beta = 0, 1, 2, 3$), give rise to eigenstates that {\it live} in both valleys simultaneously
(see below), thus giving rise to intervalley scattering.

We will now show how can we express the boundary condition matrix $\mathcal{M}$ in Eq. (\ref{eq:GeneralBC-M}) in terms 
of the generalized potential $\mathbb{V}$. We shall start by using the fact that the problem is translation 
invariant along the grain boundary direction, $\mathbf{e}_{x}$, so that we can write the eigenspinors as
\begin{eqnarray}
  \Phi(x,y) &=& \phi(y) e^{i q_{x} x} \,,
\end{eqnarray}
which allows us to rewrite Eq. (\ref{eq:GenHam}) as
\begin{eqnarray}
  v_{F} \Big( \tau_{3} \otimes \sigma_{1} q_{x} + \tau_{0} \otimes \sigma_{2} (-i \partial_{y}) + \frac{\mathbb{V}}{v_{F}} 
  \Big) \phi(y) &=& \epsilon \phi(y) \,. \nonumber \\
\end{eqnarray}
This expression can be cast as
\begin{eqnarray}
  \partial_{y} \phi(y) &=& i \hat{\mathbb{P}} \phi(y) \,,
\end{eqnarray}
where the operator $\hat{\mathbb{P}}$ reads
\begin{eqnarray}
  \hat{\mathbb{P}} &=& \frac{\tau_{0} \otimes \sigma_{2}}{v_{F}} (\epsilon \, \tau_{0} \otimes \sigma_{0} - v_{F} q_{x} 
  \, \tau_{3} \otimes \sigma_{1} - \mathbb{V}) \,.
\end{eqnarray}

Integrating the differential equation, one obtains the following relation between the two sides of the strip
\begin{eqnarray}
  \phi(W) &=& e^{i W \hat{\mathbb{P}}} \phi(0) \,,
\end{eqnarray} 
which, if we take the limit $W \, \mathbb{V} \to \mathbf{v}$ when $W\to0$,\cite{Rodrigues_GBs-CA_PRB:2012} then
becomes
\begin{eqnarray}
  \phi(0^{+}) &=& \mathcal{M} \phi(0^{-}) \,. \label{eq:phi-bc}
\end{eqnarray}
In Eq. (\ref{eq:phi-bc}) the boundary condition matrix $\mathcal{M}$ reads
\begin{eqnarray}
\mathcal{M} &=& e^{-i \frac{\tau_{0} \otimes \sigma_{2}}{v_{F}} \mathbf{v}} \,, \label{eq:BCexp}
\end{eqnarray}
with $\mathbf{v}$ reading 
\begin{eqnarray}
  \mathbf{v} &=&  \sum_{\alpha, \beta = 0}^{3} v_{\alpha \beta} \tau_{\alpha} \otimes \sigma_{\beta} \,,
\end{eqnarray}
where $v_{\alpha \beta} = W \, V_{\alpha \beta}$.

The generalized potential $\mathbb{V}$ will be hermitian if all the $v_{\alpha \beta}$ are real numbers. Time-reversal 
symmetry requires that $v_{0 2} = v_{1 2} = v_{2 2} = v_{3 0} =v_{3 1} = v_{3 3} = 0$. Therefore, $\mathbf{v}$ reads
\begin{eqnarray}
  \mathbf{v} &=& \tau_{0} \otimes \big( v_{0 0} \, \sigma_{0} + v_{0 1} \, \sigma_{1} + v_{0 3} \, \sigma_{3} \big)
  \nonumber \\ &+& \tau_{1} \otimes \big( v_{1 0} \, \sigma_{0} + v_{1 1} \, \sigma_{1} + v_{1 3} \, \sigma_{3} \big) 
  \nonumber \\ &+& \tau_{2} \otimes \big( v_{2 0} \, \sigma_{0} + v_{2 1} \, \sigma_{1} + v_{2 3} \, \sigma_{3} \big) 
  \nonumber \\ &+& v_{3 2} \, \tau_{3} \otimes \sigma_{2} \,.
\end{eqnarray}
Thus, the argument of the exponential in Eq. (\ref{eq:BCexp}) can be recast as
\begin{eqnarray}
  -i \frac{\tau_{0} \otimes \sigma_{2}}{v_{F}} \mathbf{v} = - \frac{i}{v_{F}} &\Big[& 
  \tau_{0} \otimes \big( v_{0 0} \, \sigma_{2} - i v_{0 1} \, \sigma_{3} + i v_{0 3} \, \sigma_{1} \big) 
  \nonumber \\ &+& \tau_{1} \otimes \big( v_{1 0} \, \sigma_{2} - i v_{1 1} \, \sigma_{3} + i v_{1 3} \, \sigma_{1} \big) 
  \nonumber \\ &+& \tau_{2} \otimes \big( v_{2 0} \, \sigma_{2} - i v_{2 1} \, \sigma_{3} + i v_{2 3} \, \sigma_{1} \big) 
  \nonumber \\ &+& v_{3 2} \, \tau_{3} \otimes \sigma_{0} \Big] \,,
\end{eqnarray}

In order to determine which generalized potential terms are present whenever we have a boundary condition matrix as that of Eq. 
(\ref{eq:BCmatrixP3}) we can use Lagrange-Sylvester interpolation,\cite{Horn-Johnson:1991_L-S_formula} which allows us to 
express the function of a diagonalizable matrix $A$ as
\begin{eqnarray}
f(A) &=& \sum_{i=1}^{k} f(\lambda_{i}) A_{i} \,,
\end{eqnarray}
where $\lambda_{i}$ are the eigenvalues of the matrix $A$. The matrices $A_{i}$ stand for the Frobenius 
covariants of matrix $A$.\cite{Horn-Johnson:1991_L-S_formula} These are given by 
\begin{eqnarray}
A_{i} &=& \prod_{j=1 (\neq i)}^{k} \frac{1}{\lambda_{i}-\lambda_{j}} (A-\lambda_{j} I) \,,
\end{eqnarray}
where $I$ identifies the identity matrix.

By computing $\mathbf{v} = f(\mathcal{M}) \equiv i v_{F} \log \big( \mathcal{M} \big)$, we will be able to express 
the coefficients $v_{\alpha \beta}$ as functions of $f(\xi_{1},\xi_{2},\xi_{3})$ and $g(\xi_{1},\xi_{2},\xi_{3})$ appearing in 
the expression for the boundary condition matrix $\mathcal{M}$, Eq. (\ref{eq:BCmatrixP3}). The non-zero terms of the
generalized potential originating from the 3-periodic pentagon-only grain boundary read (the principal value of the 
logarithm was taken)
\begin{subequations} \label{eq:vs-expr}
  \begin{eqnarray}
    v_{0 0} &=& \frac{1}{2} \sum_{\nu=\pm1} W_{\nu} \, , \\
    v_{0 1} &=& \frac{1}{4 \vert g \vert } \sum_{\nu=\pm1} Y_{\nu} W_{\nu} \, , \\
    v_{1 0} &=& \frac{g_{r}}{2 \vert g \vert} \sum_{\nu=\pm1} \nu \, W_{\nu} \, , \\
    v_{1 1} &=& \frac{-g_{r}}{4 \vert g \vert} \sum_{\nu=\pm1} \nu Z_{\nu} W_{\nu} \, , 
  \end{eqnarray}
\end{subequations}
while $v_{2 0} = v_{1 0} \, g_{i} / g_{r}$ and $v_{2 1} = v_{1 1} \, g_{i} / g_{r}$. All the other potential terms 
are zero: $v_{0 3} = v_{1 3} = v_{2 3} = v_{3 2} = 0$ (time-reversal symmetric); $v_{0 2} = v_{1 2} = v_{2 2} = v_{3 0} = v_{3 1} 
= v_{3 3} = 0$ (non time-reversal symmetric). Above we have used the definitions
\begin{subequations} \label{eq:vs-expr2}
  \begin{eqnarray}
    W_{\pm} &\equiv& \frac{1}{X_{\pm}} \log \bigg[ \frac{Z_{\pm}+X_{\pm}}{Z_{\pm}-X_{\pm}} \bigg] \, , \\
    X_{\pm} &\equiv& \sqrt{-4 + f_{r}^{2} + \vert g \vert^{2} \pm 2 f_{r} \vert g \vert } \, , \\
    Y_{\pm} &\equiv& \vert g \vert^{2} \pm f_{r} \vert g \vert \, , \\
    Z_{\pm} &\equiv& f_{r} \pm \vert g \vert \, ,
  \end{eqnarray}
\end{subequations}
where $\vert g \vert \equiv \sqrt{g_{r}^{2} + g_{i}^{2}}$, while $f_{r} \equiv f(\xi_{1},\xi_{2},\xi_{3})$, 
$g_{r} \equiv \Re[g(\xi_{1},\xi_{2},\xi_{3})]$ and $g_{i} \equiv \Im[g(\xi_{1},\xi_{2},\xi_{3})]$ -- see Eqs. (\ref{eq:fandg}).

We can readily conclude from the above expressions that, for a general choice of the hopping parameters of the 3-periodic
pentagon-only grain boundary, $\xi_{1}$, $\xi_{2}$ 
and $\xi_{3}$, the Dirac fermions will feel a generalized potential both containing terms that do not mix the valleys (namely, 
$v_{0 0}$ and $v_{0 1}$), and terms that do mix valleys (such as $v_{1 0}$, $v_{1 1}$, $v_{2 0}$ and $v_{2 1}$). Let us briefly examine 
the implications of the presence and absence of these terms.

Start by noting that when we force $\xi_{i} \to \xi$ the valley mixing terms vanish (i.e., $v_{1 0}, v_{1 1}, v_{2 0}, v_{2 1} \to 0$), 
and only the valley preserving terms ($v_{0 0}$ and $v_{0 1}$) are present inside the strip. As a consequence, there will 
be no intervalley scattering, just as expected: remember that when $\xi_{i} = \xi$ we recover the pentagon-only grain boundary 
which has a periodicity that does not map the Dirac points into the same $k_{x} a$, thus forbidding low-energy intervalley 
scattering.\cite{Rodrigues_GBs-CA_PRB:2012,Rodrigues_GBs-TB_JPCM:2013} This can be also concluded from the boundary condition 
matrix $\mathcal{M}$ expression when $\xi_{i} = \xi$ [given in Eq. (\ref{eq:BCmatrix-SimplePO})]: it has no off-diagonal 
(intervalley scattering) elements.

When $\xi_{i} = \xi$ there will always be an angle $\widetilde{\theta}$ with perfect transmittance, i.e. with $T=1$. 
This can be understood by noting that for this particular angle of incidence $\widetilde{\theta}$ it is possible to 
perfectly match the wave-function immediately inside the strip (at $y=-W/2^{+}$) and that immediately outside the 
strip (at $y=-W/2^{-}$) without the need to use reflected modes. As argued in Ref. \onlinecite{Rodrigues_GBs-TB_JPCM:2013},
we can see this by comparing the spinors of the modes inside the strip (Dirac modes subject to a generalized potential 
with the terms $v_{0 0}$ and $v_{0 1}$; these are non-chiral due to $v_{0 1} \neq 0$) and the spinor of the incident mode: for 
the angle $\widetilde{\theta}$ the incident mode's spinor is exactly equal to that of a positive-propagating mode inside 
the strip.

Let us now focus on the valley-mixing terms of the generalized potential $\mathbb{V}$. Both the terms $v_{1 0}$ and $v_{2 0}$ 
give rise to a shift of the energy cones along the $k_{y}$-direction (which causes a deflection of the incoming mode), 
resembling what happens when a constant gauge potential term $v_{0 2}$ is present. The latter term's eigenstates (as well 
as those of $v_{0 1}$ terms) have a well defined valley quantum number, but are non-chiral (pseudo-spin not aligned with 
momentum). Similarly, the $v_{1 0}$ and $v_{2 0}$ eigenstates are non-chiral. More importantly, and unlike what happens with 
the gauge term $v_{0 2}$ (and $v_{0 1}$), the terms $v_{1 0}$ and $v_{2 0}$ mix the two valleys, i.e. their eigenstates do not 
have a well defined valley quantum number. Similarly, we can also show that both the terms $v_{1 1}$ and $v_{2 1}$ open a gap 
in the spectrum, resembling what happens when a mass term (i.e. $v_{3 0}$) is present. However, unlike the latter, the former 
potential terms' eigenstates are both non-chiral and mix the two valleys.

The fact that (for general values of the hopping renormalizations $\xi_{i}$) there are valley mixing potential terms inside
the strip, implies that its modes do not have a well defined valley quantum number, i.e. strip eigenstates {\it live} in 
both valleys. Therefore, a wave function (living only on the valley $\nu=\pm$) incoming from $y = -\infty$, will in general 
require reflected modes (in both valleys) in order to match the wave function inside the strip. That is, in general there 
will not be an angle of perfect transmission (of low-energy carriers) at the 3-periodic pentagon-only grain boundary, i.e. 
$T = T_{\nu,\nu} + T_{-\nu,\nu} \neq 1$. Only for very particular cases, and by fine tuning the values of the hopping parameters 
at the grain boundary will perfect transmission occur.

%

\end{document}